\documentclass[preprint]{elsarticle}
\usepackage{lineno}
\modulolinenumbers[5]
\usepackage[colorlinks=true, urlcolor=blue]{hyperref}
\usepackage{listings}
\usepackage[usenames, dvipsnames]{color}
\usepackage{ulem}

\journal{NIMA}

\begin{document}

\begin{frontmatter}
\title{MCViNE -- An object oriented Monte Carlo neutron ray tracing
  simulation package}

\author[cacrcaltech,aphmscaltech,ndavornl]{Jiao Y. Y. Lin \corref{jiao}}
\ead{linjiao@ornl.gov, linjiao@caltech.edu}

\author[aphmscaltech]{Hillary L. Smith}

\author[ndavornl]{Garrett E. Granroth \corref{garrett}}
\ead{granrothge@ornl.gov}

\author[qcmornl]{Douglas~L.~Abernathy}
\author[qcmornl]{Mark D. Lumsden}
\author[qcmornl]{Barry Winn}
\author[qcmornl]{Adam A. Aczel}
\author[cacrcaltech]{Michael~Aivazis}
\author[aphmscaltech]{Brent Fultz \corref{btf}}
\ead{btf@caltech.edu}

\cortext[jiao,garrett,btf]{Corresponding author}

\address[cacrcaltech]{
  Caltech Center for Advanced Computing Research, California Institute
  of Technology
}
\address[aphmscaltech]{
  Department of Applied Physics and Materials Science, California
  Institute of Technology
}
\address[ndavornl]{
  Neutron Data Analysis and Visualization Division, Oak Ridge National Laboratory
}
\address[qcmornl]{
  Quantum Condensed Matter Division, Oak Ridge National Laboratory
}

\begin{abstract}
MCViNE (Monte-Carlo VIrtual Neutron Experiment) %\cite{MCViNE}
is an open-source
Monte Carlo (MC) neutron ray-tracing software % suite
%that provides researchers with 
% of tools 
for performing computer modeling and simulations that
mirror real neutron scattering experiments. 
We exploited the close similarity between how instrument components are designed
and operated and how such components can be modeled in software.
For example we used object oriented programming concepts
%By adopting modern software engineering practices such as using composite
%and visitor design patterns
% \cite{GammaDP} 
for representing neutron scatterers and detector systems, and 
%using 
recursive algorithms for implementing multiple scattering.
% the flexible % extensible % and sustainable architecture of 
Combining these features together in MCViNE allows one %can
%is flexible enough 
to handle sophisticated neutron
scattering problems in modern instruments, including, 
for example, 
neutron detection by complex detector systems,
%like those of ARCS \cite{abernathy2012design} and SEQUOIA \cite{Granroth2006,Granroth2010} at SNS,  
% as well as
and single and multiple scattering events in a variety of samples and
sample environments.
In addition, MCViNE can use
%take advantage of
simulation components from 
%in
linear-chain-based MC ray tracing packages which facilitates porting instrument models from those codes.
% such as McStas \cite{lefmann1999mcstas, willendrup2004mcstas}, 
% which is 
%widely used in instrument design and optimization,as well as
Furthermore it allows for  components written solely in Python, which % facilitates and
expedites prototyping of new components.
 %NumPy-based components that 
%\cite{numpy}
%This 
%make prototypes useful and easy to develop.
% easy to develop and very useful in prototyping.
These developments
% , combined with improvements in computing
% hardware, 
have enabled 
%us to carry out
detailed %MC ray tracing 
simulations %that capture unprecedented details of
of neutron scattering experiments, with non-trivial samples,
%in %some 
for time-of-flight
inelastic instruments at the Spallation Neutron Source.
Examples of such simulations % of experiments at these instruments
for powder and single-crystal samples with various scattering
kernels, including kernels for phonon and magnon scattering, are presented.
With simulations that closely reproduce experimental results, 
scattering mechanisms can be turned on and off to determine how they
contribute to the measured scattering intensities, improving our
understanding of the underlying physics.
% of neutron scattering XXX
\end{abstract}


\end{frontmatter}
%
%
% \linenumbers

\lstset{
  language=Python, 
  basicstyle=\ttfamily\footnotesize, 
  keywordstyle=\color{blue},
  commentstyle=\color{Gray},
  stringstyle=\color{red},
  showstringspaces=false,
  identifierstyle=\color{black}
}

\section{Introduction}
\label{intro}
Data analysis for neutron time-of-flight spectroscopy has predominantly been 
a conversion of raw data to intensities with instrument independent
units; namely $S(\textbf{Q},\omega)$ or $S(\textbf{Q},E)$,
where $S$ is the dynamical structure factor,
and $\textbf{Q}$, $\omega$, $E$
% $E=\hbar\omega$
are the
wave vector, frequency, and energy. % of the excitation,
% respectively.
This approach has been the basis for many studies,
but %it has become clear that in
some cases 
%this simple ``reduction'' will not suffice
require a more % detailed reduction scheme
elaborate analysis scheme
(see, for example, \cite{
  HugouvieuxPRB2007, walters2009effect,
  delaire2011phonon, weber2012electron,
  aczel2012quantum, 
  StevenHahnPRB2013YFeO3, 
  PlumbPRB2014Sr2CuO2Cl2, 
  LiChenPRL2014, linjyy2014UNprb}).

%As neutron instruments gain more flux and capabilities, measurements
%of complex samples have become more common.  
More specifically, traditional 
%With higher count rates, 
%smaller effects become statistically significant, and it is important
%to develop methods to interpret them.
reduction assumes single scattering,
that the resolution function can be modeled by a simple function,
that the scattering process of interest can be readily separated
from other scattering processes in the system,
and that scattering from the sample environment is negligible
or easily subtracted.
Monte Carlo ray-tracing can provide a straightforward way
to perform %reduction and
analysis in cases where some,
or all, of the above assumptions are not valid.
Specifically it can be used to investigate complex samples
whether they have multiple scattering,
multiple modes of scattering,
i.e. magnetic and lattice vibrations,
or consist of a conglomeration of scatterers like a sample and sample environment.
Additionally as the instrument is modeled component by component,
instrument resolution is inherently included in the simulation.

As the promise of Monte Carlo ray tracing was clear,  it has been used in codes that handle a subset of these complexities.
MCS\cite{bischoffmsc} and MSCAT\cite{copley1986improved} 
were early packages that used Monte Carlo methods
to compute multiple scattering.
Later,
several
general-purpose Monte Carlo neutron instrument simulation packages were developed and optimized 
to help design neutron instruments,
% such as
namely 
McStas \cite{lefmann1999mcstas, willendrup2004mcstas},
VITESS \cite{vitess2002}, IDEAS \cite{LeeWTideas}, and NISP \cite{NISPspie1999}.
Less-general-purpose MC programs exist in some popular 
neutron data  analysis software packages, 
including DISCUS \cite{proffen1997discus, proffen1999discus, DISCUS}
and RESTRAX \cite{vsaroun1997restrax, RESTRAX}.
In comparison,
relatively few studies
\cite{HugouvieuxPhysicaB2004, Udby2011analysing,
  tregennaJNR2008reduction, boin2011validation,
  boin2012nxs}
have employed full-fledged Monte Carlo ray tracing to help analyze
experimental results from neutron scattering measurements,
while there are growing efforts to perform
virtual neutron experiments using MC ray tracing
\cite{lefmann2008virtual,  
farhi2009virtual,
FarhiCSFN2011,
willendrup2011complexoptics,
farhi2014advanced}.
MCViNE\cite{MCViNE}, an open source software, is designed to easily allow 
complex studies (see \cite{linjyy2014UNprb} for an example)
and therefore should accelerate the use of 
Monte Carlo ray tracing in the analysis of experimental data.

The improvements in computing over the last decades
include hardware developments,
the emergence of object-oriented languages C++ and Python, and
advances in software design \cite{GammaDP}. 
The amazing increases in hardware capabilities
now allows detailed simulations to run in a reasonable time.
The impact of modern software engineering practice on MC simulation codes is still progressing.
%For MC simulations of neutron scattering experiments,
%the impact of these improvements 
%on modern software engineering practice,
%especially object-oriented language features, is still being realized.
MCViNE was developed, as part of the DANSE project \cite{DANSE}, using such software practices.
%and its goal is to simplify the set-up of  Monte Carlo ray tracing simulations of
%neutron experiments, making it possible for non-experts to run 
%simulations of non-trivial samples and sample assemblies
%and analyze neutron data with more quantitative accuracy.
Therefore it uses object oriented programing (OOP) concepts to  represent instrument and sample components,
 following the hierarchical and modular nature  of the neutron instruments, detector systems, and sample environments. 
%inspired a hierarchical and modular 
%representation of neutron scatterers in MCViNE,
%going beyond the traditional
%approach of simulating a linear sequence of neutron components.
%Consequently 
%MCViNE employs object-oriented programming (OOP), to provide a structure that better mirrors the structure of the neutron instrument hardware,
%instead of the 
This use of hierarchies from inside an instrument component or sample kernel up through the total instrument goes beyond the imperative programming paradigm, which is popular in other codes.\footnote{A brief discussion is available in the supplemental material.}
%imperative programing paradigm that is  popular in other
% software.
By taking advantage of OOP design patterns
as well as recursive algorithms, 
the MCViNE architecture supports
an easily extensible library for scattering kernels
suitable for both samples and detector systems.
This approach allows for maximum flexibility, extensibility, and reuse of
scatterer arrangements, geometrical shapes,
and scattering mechanisms, and hence improves the sustainability of the
software.

% ??? rewrite ???
This paper contains a description of 
the MCViNE software framework.
%including %abstract inaterfaces, 
%some key abstractions,
%and algorithms
%built on top of those abstractions. % abstract interfaces.
Section~\ref{mcvine} presents 
the challenges in simulating experiments carried out 
in modern neutron instrumentation,
and an overview of the software engineering design and 
main software components and algorithms.
% an overview of the
% architecture of the MCViNE software framework.
% and explains how the software constructs in MCViNE allow full
% extensibility for scattering mechanisms,
% especially the hierarchical representation of neutron scatterers. 
% Algorithms for multiple scattering and neutron
% detection in a detector system are then presented.
Examples of MCViNE simulations for powder and single-crystal
experiments are
presented in Section~\ref{examples}.
Concluding remarks follow in Section~\ref{conclusions}.

\section{Simulation of Spectra Measured at Modern Neutron Spectrometers with MCViNE}
\label{mcvine}

\subsection{Challenges for simulating neutron scattering experiments}
% Neutron spectroscopy is a great technology and has a long history
% of successful use in studying material structure and dynamics.
% The technology has undergone tremendous growth.
% When creating new instruments, Monte Carlo simulatons
% are especially useful in studying performance of guides
% and design and optimization of instruments.
% Although analytical calculations may be used to obtain the general
% behavior of the neutron instruments, MC simulations are 
% useful to confirm the intuitions and obtain optimal 
% parameters for the design.
% Typical instruments consist of a linear chain of neutron optical components,
% starting from moderators through guides etc to the sample position,
% where neutrons are scattered by the sample, and
% then intercepted by detectors.

Monte Carlo (MC) ray tracing simulations of neutron scattering spectrometers
with support of multiple scattering
were performed from 1970s using
MCS\cite{bischoffmsc} and 
MSCAT\cite{copley1986improved}.
They were used to
%, for example,
% analyze the performance of guides \cite{copley1993},
% and 
%study 
understand the effects of multiple scattering on the measured spectra
\cite{wuttke2000}.
In the 1990s, with increasing needs of simulating neutron
instruments for the purpose of instrument design and optimization,
several MC neutron ray tracing packages emerged, 
including McStas\cite{lefmann1999mcstas, willendrup2004mcstas},
Vitess\cite{vitess2002}, Ideas\cite{LeeWTideas},
and NISP \cite{NISPspie1999}.
Simulations performed with these packages
provide not only independent checks for
analytical calculations of instrument performance,
but also can be used 
to obtain optimal parameters for instrument design
(see, for example, 
\cite{SorenKynde2014ESS,Klinkby2014ESS,Bertelsen2013387ESS,
Prokhnenko2014VITESSMultPurpose,
Houben2012124,Skoulatos2011100HZB,Zunbeltz2010NEAT,
Alianelli2004231,Wildes2002McStasFocusingGuide, Granroth2003}).
Unlike MCS and MSCAT, 
most of these newer MC software packages 
(with a notable exception in NISP) treat simulation
of a neutron instrument as a linear chain of neutron
optical components, each of which modify %ing 
some set of neutron beam characteristics such as spatial divergence
and energy distribution.
This linear approach greatly improved the computational efficiency
and simplified the %programming
coding of instrument simulations.
Such an approach is adequate because the physical formation 
of a neutron scattering instrument is linear, 
and the underlying
hypothesis that neutrons at the downstream components are rarely
scattered back to the upstream components holds up
well in most instrument configurations.
As a result, these software packages, 
especially McStas and Vitess,
are making significant impacts for neutron instrument design.
%still in wide-spread use nowadays.

% The linear chain formation of the neutron instruments make
% it natural to use a linear chain of software components
% to simulate the neutron instruments.
% This simplification is conceptually simple 
% and easy to implement in practice.
% McStas makes the software especially easy for
% others to contribute by allowing authors to write
% a small component using C and wrap it with a few
% lines of meta language code.
% And this is one reason why such MC simulation packages
% are successful earlier.
For Monte Carlo neutron ray tracing simulations
to be useful for interpretation of 
neutron scattering spectra,
it is necessary to include 
detector systems and samples/sample environments,
for which the physical arrangements are often
non-linear.
The detector systems 
in modern neutron scattering instrumentation
display modularity,
repetition, and sometimes hierarchical organization,
reducing the engineering difficulties in
manufacturing, testing, and validation.
For example,
the four direct geometry time-of-flight spectrometers at 
Spallation Neutron Source (SNS) \cite{mason2000sns}, 
ARCS \cite{abernathy2012design}, 
SEQUOIA \cite{Granroth2006,Granroth2010},
CNCS\cite{Ehlers2011},
and
HYSPEC\cite{Barry2015}
share some instrument elements\cite{mattstone2014compareDGTOF}:
the Fermi chopper slit packages for 
the ARCS and SEQUOIA
instruments are interchangeable, and the detector 
systems of all four instruments use the same
so-called 8-packs \cite{Riedel2012}, 
each of which is a detector pack consisting of
eight $^3$He linear position sensitive detector tubes (LPSD).
The 8-packs are arranged 
in a vertically-oriented cylindrical geometry 
around the sample position,
forming a hierarchical organization 
of pixels, tubes, packs, and detector rows.
% (three rows for ARCS and SEQUOIA
% and one row for CNCS
% and HYSPEC
% \footnote{
A simplified illustration of the detector hierarchy
can be found in Figure~\ref{fig-composite-examples}(c).

Samples, sample holders, and sample environments
are 
constituents of a ``sample assembly'',
the term we use to refer to the collection of neutron
scatterers near the sample position.
A sample-assembly example is illustrated in 
Figure~\ref{fig-composite-examples}(b).
A linear representation
faces challenges in simulations of a sample assembly
because it does not match physical
arrangements of a sample or
samples with associated sample environment components,
and because
neutrons can often scatter back and forth among the 
constituents. % in a sample assembly.
A good example of complex sample environments
is the MICAS furnace \cite{jennifer2015MICAS}: % for the ARCS instrument 
the incident neutron beam intersects heating elements % (rods),
% and a choice of at maximum 5 vanadium heat shields
and up to 8 heat shields, all thin vertical hollow cylinders,
in addition to the sample.
The sample itself can be challenging too.
For example, ``single crystal'' superconductor samples for inelastic
neutron scattering experiments often consist of co-aligned
arrays of small crystals mounted on an aluminum plate
\cite{friemel2013coaligning}.
Another major difficulty stems from the fact that
a variety of different scattering mechanisms may be
present in the combination of sample, sample holder,
and sample environment.

% Samples and sample environments are more complex than before.
% One important thing at sample is that multiple scattering 
% can be important.
% It could be the multi-layer furnace.
% It could be just the sample is large.
% It could be unconventional? arrangement of the samples.
% All these are not easily dealt with if we have only
% a linear simulation package.
% And it has many different scattering mechanisms.
% How to have a flexible framework to allow all these?
% Some samples may have modular and even hierarchical structure?

% sample and sample environment makes it absolutely
% necessary to go beyond linear approach.
% modular and hierarchical structure prevail
% in sample and sample environment.
% sample kernels --> different scattering mechanisms.

% Growing community of neutron scattering science and
% growing complexity in neutron scattering experiments
% for more sophisticated instrumentation and sample
% call for better analysis and better simulations.

% Nowadays, we want to simulate not just the neutron beam, 
% but also the spectrum measured.
% Also in the modern instrumentations, detector systems
% are complex and the sample environments are complicated
% too.

% Hierarchical structure and modular structure are ubiquitous.

% He3 detector systems are prevalent before, 
% but became expensive and hard to get.
% Other alternatives come in. 
% But we would imagine they are still modular and 
% probably with hierarchy.

\subsection{General capabilities of MCViNE}
MCViNE\cite{MCViNE} is a general purpose neutron ray tracing package that 
combines the two different approaches taken by MSCAT-like packages
and McStas-like packages.
At the instrument level, it allows users to create a simulation application
as a linear chain of neutron components, each fully configurable by
its type and corresponding geometrical and physical
properties using either command line options
or an xml-based configuration file.
At the component level, two general components exist with
support of hierarchical representation for 
the sample assembly and the detector system.
A detector system can be specified with an xml
file describing its hierarchy,
while a sample assembly is specified
by a collection of xml files,
one for the geometrical organization
of the constituent scatterers,
and others describing the scattering mechanisms for 
them.
Multiple-scattering among scatterers in a
sample assembly can be turned on and off 
by a command line option.

\subsection{Architecture Overview}
\label{arch}

% There are two classes of Monte Carlo (MC) neutron ray tracing software packages. 
% The first class takes a linear approach, 
% and is in widespread use because it has proved critical for instrument
% design (see, for example,
% ).
% It includes packages such as 
% McStas\cite{lefmann1999mcstas, willendrup2004mcstas}, Vitess\cite{vitess2002}, and Ideas\cite{LeeWTideas}.
% The second class has inherent support for multiple scattering
% including MCS\cite{bischoffmsc},
% MSCAT\cite{copley1986improved}, and NISP\cite{NISPspie1999}.
% For example, MSCAT
% is not implemented using the linear execution structure, and
% handles multiple scattering between sample and sample container
% in a more generic way than linear programs.
% However, this package was written
% for a few specific cases on specific instrumentation, %using Fortran 77,
% creating difficulties in extending it to 
% more sophisticated instrument configurations and sample assemblies.

MCViNE is implemented in C++ and Python.
%At the 
The C++ portion
% level, 
contains fundamental mathematics, the neutron ray tracing mechanism, the mechanism to include components from other MC ray tracing packages,
%was implemented  with 
and support of multiple scattering and composite scattering kernels.  
These various functions of the C++ code are implemented in two layers, which are discussed  in the supplementary information.
There are two aspects of the C++ usage that bear more discussion.  
The components from McStas  are included in the C++ layer
so that they can be used as individuals rather than as a whole compiled instrument.
Also, the combination of the object oriented features of C++ and the speed of the compiled code make this the appropriate place to implement the composite kernel and scattering kernels. 
%3) Components that use the notion of composite scatterer and
%scattering kernels. 
Currently two such components exist, one for
sample assemblies and one for detector systems.
This unique feature of MCViNE, the ``composite scatterer'',
will be explained in detail in section \ref{composite-scatterer}.

A Python layer on top of the two C++ layers %however, 
allows construction of a component
chain similar to
% the first class of
McStas-like
neutron ray tracing packages,  provides the interface to the C++ components to include in that chain, and  allows for introduction of components that are completely written in Python.
This last feature makes it extremely easy to create
simple neutron components, and to create prototypes of more sophisticated
neutron components.
% One effect of using Python components is that neutron packets are
% passed through the component chain in a group, to reduce the overhead
% of calling C/C++ functions from Python.
% Each group of neutrons is represented by an array.

\begin{figure}
\centering
\includegraphics[width=8.3cm,clip]{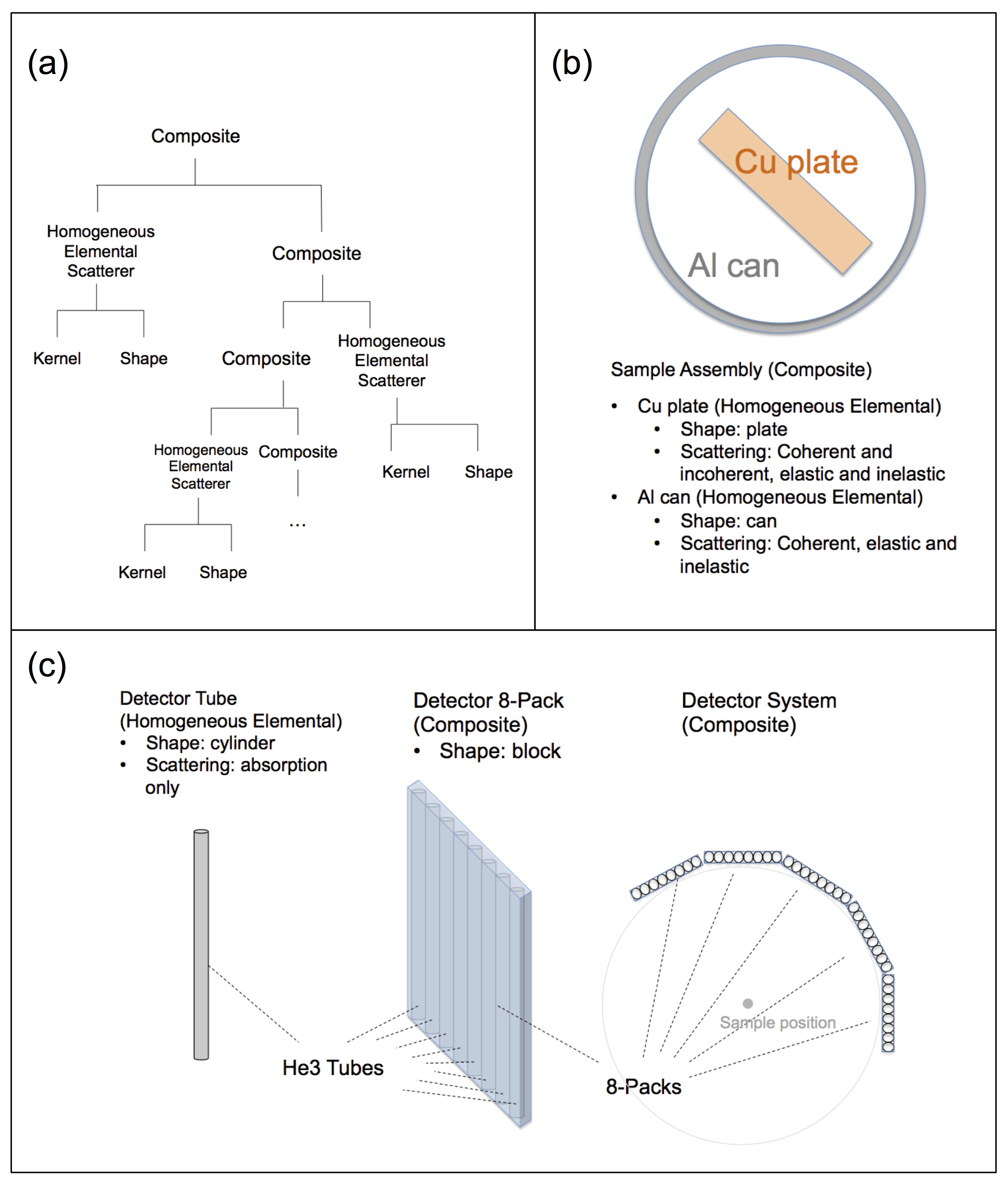}
\caption{Concepts in scattering composites: composite scatterer,
  homogeneous scatterer, shape, scattering kernels. (a) an abstract hierarchy
  of an abitrary scattering composite. (b) and (c) are
  concrete examples of such hierarchies. (b) top view of a sample
  assembly consisting of an aluminum can and a copper plate. (c)
  a detector system -- how it is constructed from an elemental
  scatterer (detector tube) in a three-level hierarchy.}
\label{fig-composite-examples} 
\end{figure}

\subsection{Composite Scatterer}
This section starts with
basic concepts with regard to the ``composite scatterer'',
followed by an introduction to
the essential object-oriented software designs centered around it,
and finishes with examples of ray-tracing algorithms in
sample assemblies and detector systems, enabled by these designs
and needed in simulation of neutron scattering experiments
in modern spectrometers.

\label{composite-scatterer}

\subsubsection{Concepts}
In MCViNE, a ``composite neutron scatterer'' represents
a group of physical objects, for example, 
a powder sample in an aluminum can,
a single crystal sample surrounded by a furnace,
or a detector system.
An ``elemental scatterer'' is a scatterer without constituent scatterers.
A ``homogeneous scatterer'' is one kind of elemental neutron scatterer,
whose scattering function is homogeneous within its volume.
The scattering properties are modeled using
one ``scattering kernel''
or a combination of several ``scattering kernels'',
each of which represents one scattering mechanism,
such as incoherent one-phonon nuclear scattering
or coherent magnetic scattering.

These concepts are illustrated in Figure~\ref{fig-composite-examples}.
Figure~\ref{fig-composite-examples}(a) is an example of an abstract
hierarchy of a composite scatterer in MCViNE. 
In principle, the hierarchy in MCViNE can be arbitrarily deep.
In practice, the depth of the hierarchy is limited by factors such as
computing resources available for the simulations, % and language
and compiler limitations.
Figure~\ref{fig-composite-examples}(b) depicts a sample assembly
that is a composite of two-level hierarchy, in which the bottom level
consists of two homogeneous elemental scatterers: one aluminum can
and one copper plate sample.
Figure~\ref{fig-composite-examples}(c) represents a detector system 
consisting of $^3$He eight-packs that form roughly a cylindrical
arrangement around the sample position.
The detector system is represented in MCViNE in a three-level
hierarchy: at the bottom level is the $^3$He detector tube; at the
middle level, 8 such tubes construct an 8-pack; at the top-level, the
detector system consists of a collection of 8-packs.
Such hierarchical representations allow MCViNE to model the physical
reality closely.

\subsubsection{Object Oriented Designs}
These concepts lead to one major design decision made in the MCViNE
project:
employment of the 
``composite design pattern'', %\footnote{Composite pattern can be
which is used to describe the part-whole relationship and to represent a hierarchy, and allows clients to treat composites and their constituents in a uniform way, %}
and the ``visitor design pattern'',
% \footnote{Visitor pattern 
which allows separation of operations
from the objects to be operated on, so that new operations can be
added without touching these objects %.}
% design patterns of
%of object-oriented programming
\cite{GammaDP}.
By using these design patterns for scattering composites, 
we can unify the programming
interfaces to the operations on both the composites and the individual
elemental objects.
Composite and visitor patterns are used in three major aspects of the MCViNE
neutron scattering model: the neutron scatterers,
the geometric shapes of scatterers, and the scattering kernels.

\textbf{Neutron scatterers}. By using the composite pattern,
algorithms for multiple scattering can be consolidated
in one implementation.
Scattering from a composite neutron scatterer
starts with a determination of which constituent
intersects the incident neutron
ray, and then delegates the scattering assessment
to that particular constituent, 
which could be a composite itself that requires another delegation for
scattering.
The hierarchical representation of neutron scatterers and this recursive algorithm
work for both samples and detector systems, 
and can improve computing efficiency and code maintenance.

\textbf{Geometric shapes}
Using constructive solid geometry (CSG)
(see for example \cite{GhaliIntroToGeometricComputing}),
composite shapes are
constructed from basic shapes such as cylinders and blocks,
and composites by
using operations such as union, intersection and difference.
Ray-tracing through shapes is therefore simplified as visitor
methods of the primitive shapes and the binary shape operations.

\textbf{Scattering kernels}
%Because the scattering kernel can be a composite,
Composite scattering kernels make it easy for the Monte Carlo algorithm
to sample a total $S({\bf Q}, \omega)$
consisting of both slowly-varying regions and
regions containing sharp features, such as
diffraction and coherent phonon scattering, 
by allowing users to combine different kernels
such as incoherent and coherent kernels in a 
kernel composite.
% Scattering kernels of similar nature can be grouped into a
% composite, and this makes it convenient to apply
In addition, users can organize scattering kernels into groups;
this makes it easy to apply importance sampling 
(assign different weights to different kernels or kernel groups) 
in simulations.
%different kernels, either composite or elemental.

A ``scattering kernel'' in MCViNE is conceptually
different from sample components in linear MC neutron ray tracing packages.
A scattering kernel in MCViNE is an abstraction of the scattering
mechanisms such as diffraction, nuclear scattering by phonons, 
and magnetic scattering by spin waves.
It does not include the sample geometry but only the
scattering physics.
A sample component in earlier packages, on the other hand, includes both the
geometry and the physics in one programming unit.
By separating the implementation of ``scatterer'', ``shape'', and
``scattering kernel'', 
a sample in MCViNE can consist of a combination of
scattering kernels.
Furthermore, the scattering kernel library in the MCViNE framework
can be extended without affecting the logic of geometric ray tracing,
which is implemented in ``shape'' and ``scatterer''.
For example, for isotropic scattering, 
scattering kernels taking a histogram form of $S(Q,\omega)$ can be
supported,
as well as phonon scattering kernels taking phonon energies and
polarizations as inputs (examples are given in Section \ref{calib-v}
and \ref{calib-al}).
A scattering kernel conveniently 
taking the analytical form of a dispersion 
can be used
(see \cite{linjyy2014UNprb} for an example)
to improve convergence (effective when combined with other scattering
kernels), and to avoid unnecessary broadenings resulted from
approximating $S(Q,\omega)$ using a histogram.
Inelastic and elastic scattering kernels for single crystals
can also be developed to simulate single crystal
experiments.

\subsubsection{Algorithms}
In this section, the general 
ray tracing procedure is first described briefly,
and then implementation of multiple scattering
and ray tracing in detector systems are presented as algorithm
examples that 
benefit from the conceptual analysis and the software design.

\noindent
\textbf{Ray Tracing}

In components such as sample assemblies
or detector systems, neutron scatterers are
represented by a hierarchy of objects 
with shapes and scattering mechanisms.
Ray tracing of a neutron happens by first 
investigating which one of the neutron scatterers
at the top level of the scatterer composite hierarchy
intercepts the neutron.
This is done by
computing the intersections of the forward
ray of the neutron and all the shapes
of the top level constituents.
A random selection might be necessary if
multiple top-level objects intercept the neutron.
After the top-level neutron scatterer is identified,
the neutron is propagated to the front surface of the scatterer
if necessary (not necessary if the neutron is already inside 
the scatterer) with appropriate attenuation,
and then the ray tracing algorithm recurses
into itself if the scatterer is  a composite.
Otherwise a point in the forward path of the neutron
inside the scatterer will be 
randomly picked,
and the neutron is propagated to that point with attenuation.
At this point a scattering (or an absorption) mechanism
of the scatterer is randomly picked,
and the neutron will be either scattered 
with its probability adjusted,
or be absorbed.

%\paragraph{\label{app:MS_descript} Multiple Scattering} %******
\noindent
\textbf{Multiple Scattering}

\begin{figure}
\centering
\includegraphics[width=8cm,clip]{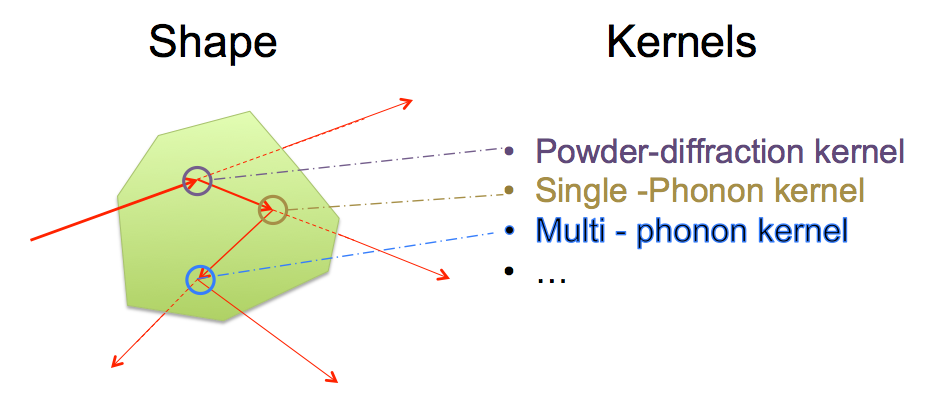}
\caption{An example of multiple scattering within one scatterer.
  The incident neutron was
  scattered three times by three different scattering kernels.
  % by a powder diffraction kernel, a
  %single-phonon kernel, and a multi-phonon kernel, respectively. 
  % To improve simulation efficiency, 
  At each scattering point, the
  original neutron is also propagated out of the scatterer with
  proper attenuation. 
  Red arrows are paths of neutron propagation. 
  Circles highlight the location of scattering. 
  Different scattering events are coded using different
  colors. 
}
% how about elemental scatterer?
\label{fig-multiple-scattering} 
\end{figure}

Multiple-scattering (MS) is naturally supported in MCViNE scattering
composites, implemented with a recursive algorithm.
For the purpose of this discussion,
we differentiate between two types of
multiple scattering: single-scatterer multiple scattering (SSMS) to describe
the multiple scattering within one neutron
scatterer and multiple-scatterer multiple scattering (MSMS) to describe 
multiple scattering among neutron scatterers.

Figure~\ref{fig-multiple-scattering} shows an example of SSMS
in which a neutron gets scattered three times before
exiting a scatterer. 
Each time a neutron is scattered inside a
homogeneous scatterer, the original incident neutron packet is split 
into two neutron packets for computational efficiency.
One neutron packet is propagated through the scatterer with its probability
lowered by attenuation, while the other is scattered by one of the
scattering kernels, chosen by a random selection,
at a point also randomly selected along the forward path of the
incident neutron.
This splitting process repeats for the scattered neutron several
times, as illustrated in Figure~\ref{fig-multiple-scattering},
until either
the neutron probability is lower than a pre-selected limit, 
or the maximum order of multiple scattering is reached. 
At this point, the neutron that is still inside the scatterer is
allowed to propagate out, 
with its probability attenuated appropriately. 

\begin{figure}
\centering
\includegraphics[width=6.5cm,clip]{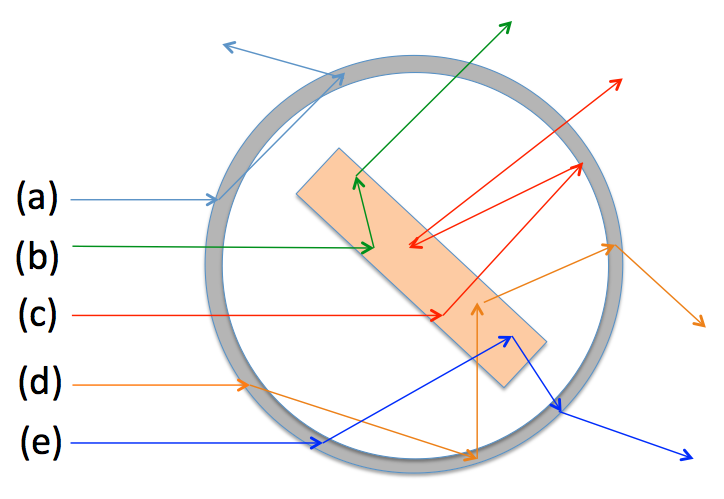}
\caption{An example of multiple scattering in a concentric sample assembly.
  Five (out of infinite) possible multiple scattering paths 
  are illustrated.
}
\label{fig-multiple-scattering-sampleassembly} 
\end{figure}

Figure~\ref{fig-multiple-scattering-sampleassembly} shows an example of multiple
scattering in a sample assembly consisting of a ``concentric''
arrangement of a sample and a sample environment.
Only five of an infinite number of possible multiple scattering paths are illustrated, and the
splitting processes are not shown.% to 
% ensure the diagram conveys the correct idea.
%avoid cluttering.

The multiple scattering algorithm of MCViNE is generic, and it can 
handle more complex sample assemblies, such as
non-concentric arrangements, and mixing of non-concentric
and concentric arrangements. 
More details about the MS algorithm of MCVINE can be found in the
supplemental material for this paper.

Sample components in some linear-chain-based MC ray tracing
packages can support SSMS,
but they do not have abstractions similar to composite scatterer or scattering
kernel. 
As a result, the SSMS
algorithm must be duplicated in 
these sample components, 
while in MCViNE, implementations of new scattering kernels can be added without
reimplementing the multiple scattering algorithm. % again and again.
McStas supports MSMS partially for a concentric sample assembly
by, for example, adding a second outer cylinder into the simulation
component  chain \cite{mcstascomponentmanual}.
Path (e) of Figure~\ref{fig-multiple-scattering-sampleassembly} is
included this way, but
path (c) of
Figure~\ref{fig-multiple-scattering-sampleassembly}
is not.
Other ways of simulating MSMS may be possible with McStas,
but these would require significantly more effort on the user's part.

The multiple scattering algorithm of MCViNE is more comparable to that of
MSCAT\cite{copley1986improved} in which the sample and the sample
environment are treated together in the multiple scattering loop, offering
complete treatment of SSMS and MSMS.
However, MCViNE allows a straightforward increase
in complexity of the
sample and sample environment interactions 
by using the recursive MS algorithm enabled by OOP.
MSCAT by default only allows for a few specific arrangements.
MSCAT and some McStas components contain the ability to 
appropriately transform the angular scattered distribution off of the
sample so as to only simulate neutrons that will impact the detector,
such optimization is not yet available in MCViNE.

\noindent
\textbf{Ray tracing in a detector system}

Ray tracing of a neutron through a sophisticated detector system 
in instruments such as ARCS and SEQUOIA,
where flat detector packs are arranged % to form
in an approximately cylindrical arrangement,
illustrates another strength
of the software design of MCViNE.
MCViNE takes advantage of the hierarchical
representation for neutron scatterers, only in this case the
elemental homogeneous neutron scatterer is the $^3$He detector tube that intercepts
neutrons and records them.
MCViNE reuses the code for ray tracing in a composite scatterer 
for simulating $^3$He detector systems,
and the new code needed is a
scattering kernel for the $^3$He material that takes into account 
gas absorption.
The ray tracing through a cylinder takes care of the % charge division,
parallax %and border
effect of the detector tube.
When a neutron is sent to a detector system shown in
Figure~\ref{fig-composite-examples}(c), 
for example, the
generic ray tracing algorithm for composite scatterers first checks
whether 
the top level composite scatterer 
% having a ``coarse-grained'' hollow-cylinder-like shape 
is penetrated by the neutron.
If so, all constituents of the composite scatterer, i.e.
the detector packs, are examined to determine which of them intercepts the
neutron.
Unless a neutron traverses a gap between detector packs, the detector
pack is identified and then
its constituents, the 8 detector tubes, are examined for neutron detection.
The path of a neutron through the detector
tube is then computed by ray tracing of the neutron 
through a cylinder, and a MC sampling 
picks a point in the path for the neutron to be absorbed.
The position in the tube is used to calculate a unique pixel
identifier known as the Pixel ID, according to the scheme readable by
the Mantid software framework
\cite{arnold2014mantid,  taylor2012mantid}.
Additionally, the appropriate weighting multiplier for neutron
probability (computed from
absorption probability depending on the $^3$He pressure and the length of
the neutron path through the tube) and the time-of-flight are computed for the neutron
% according to the absorption  
to be
recorded as a detector event in a ``virtual detector electronics
device''.
Our hierarchical approach to detectors allows for the addition of 
%as much
more details, % as necessary,
including details of the charge cloud and the
wire if deemed necessary.

\section{Examples}
\label{examples}

This section presents examples of MCViNE simulations performed for
experiments on 
the ARCS, SEQUOIA, and HYSPEC instruments at SNS.
These simulations were performed on one of
the SNS data analysis clusters. 
A typical analysis cluster has 64 Intel or AMD CPU cores
at $\sim$3GHz, and simulations usually run in parallel
using 10 cores.
Every simulation consists of 4 steps:

\textbf{Beam simulation.}  The incident beams on the sample for ARCS,
SEQUOIA, and HYSPEC instruments were simulated. 
The MCViNE simulation
models were derived from the McStas\cite{lefmann1999mcstas}
instrument definitions used in the design phase of the
instruments\cite{Granroth2003,Granroth2006,granroth2007fast,hyspecmcstasmodel}.
In this work, $10^9$ neutron events emitted from the moderator were
included in all the simulations, 
and the neutrons at the sample position were saved and reused in the
next step.
A typical parallel run took about one hour.

\textbf{Sample scattering.} The neutron packets saved in the previous
step were sent to a SampleAssembly component.
Typically $10^8$ neutron packets were simulated in a run
for any powder sample presented in this section, 
and it took minutes to hours to finish, 
depending on the complexity of the sample assembly.
The scattered
neutrons were then saved.

\textbf{Detector interception.} Each neutron scattered by the sample
was processed by a DetectorSystem component and an event was recorded
in an event-mode NeXus file if it intercepted a detector tube. 
In ray-tracing through a detector system, the detector tube in which
the event was detected was located using a hierarchical set
of detector tubes as described previously, while the exact location and
time-of-flight were determined by a MC selection. 
The running time of this step depends on the number
of neutrons hitting the detectors, which in turn depends on
the number of neutron packets simulated in the previous step,
as well as the user choice of multiple-scattering.
Typically, $10^8$ neutron packets took half an hour.

\textbf{Reduction.} The NeXus data generated in the previous step were
reduced using Mantid\cite{arnold2014mantid, taylor2012mantid}. 
The only difference between the simulated and measured data is that
the intensities in the simulations are computed as the sum of the
probabilities of all packets arriving in the bin of interest, while
those in the data are total event counts. 
The data reduction workflow is therefore identical, and uses the same code
base for both the simulated and the measured data.
{This step typically took 10s of minutes to finish for a powder
measurement.}

The goal of this four-step simulation process was to reproduce the
experiment (including data reduction) by a simulation of high
fidelity. 

This 4-step simulation workflow produces output files such as 
the simulated scattered neutrons,
the simulated event-mode NeXus file,
and the reduced $I(Q, E)$ file.

The simulation examples here make use of the following scattering
kernels that are described briefly in the Supplemental Material:
\begin{itemize}
\item incoherent elastic scattering
\item coherent elastic scattering from powder sample
\item incoherent inelastic single-phonon scattering 
\item coherent inelastic single-phonon scattering from a powder sample
\item multi-phonon scattering
\item scattering from a dispersion surface where the dispersion
  relation and the dynamical structure factor are described by
  analytical functions of momentum transfer vector $\textbf{Q}$
\end{itemize}

Three examples are presented in the following subsections,
two modeling vibrational excitations,
and one modeling magnetic excitations.
The simplest sample, a vanadium plate, is presented first in
Subsection \ref{calib-v}, where only incoherent scattering kernels are
used.
Coherent phonon scattering kernels for powder samples
are introduced for an aluminum
sample in Subsection \ref{calib-al}.
Both Subsection \ref{calib-v} and \ref{calib-al} demonstrate the ability to 
easily turn on and off different scattering mechanisms, 
allowing researchers to gain a better understanding of their contributions.
Subsection \ref{kvo} presents a simulation 
of a measurement on a single-crystal K$_2$V$_3$O$_8$ sample,
for which a dispersion-surface scattering kernel is used. 

In addition to these three examples, 
MCViNE simulations of Uranium Nitride (UN) measurements performed on 
the ARCS and SEQUOIA instruments provide an excellent example of
the capabilities presented here \cite{linjyy2014UNprb}.
In this case, the UN sample exhibited particularly strong multiple
scattering due to its size.  
MCViNE simulations were able to reproduce well the multiple
scattering, identified weak
scattering from accoustic phonons, and 
showed that the binary solid model is a good explanation of the
temperature-dependent broadening of the modes of 
the quantum harmonic oscillator \cite{linjyy2014UNprb}.

\subsection{Vanadium}
\label{calib-v}

\begin{figure}
\centering
\includegraphics[width=7cm,clip]{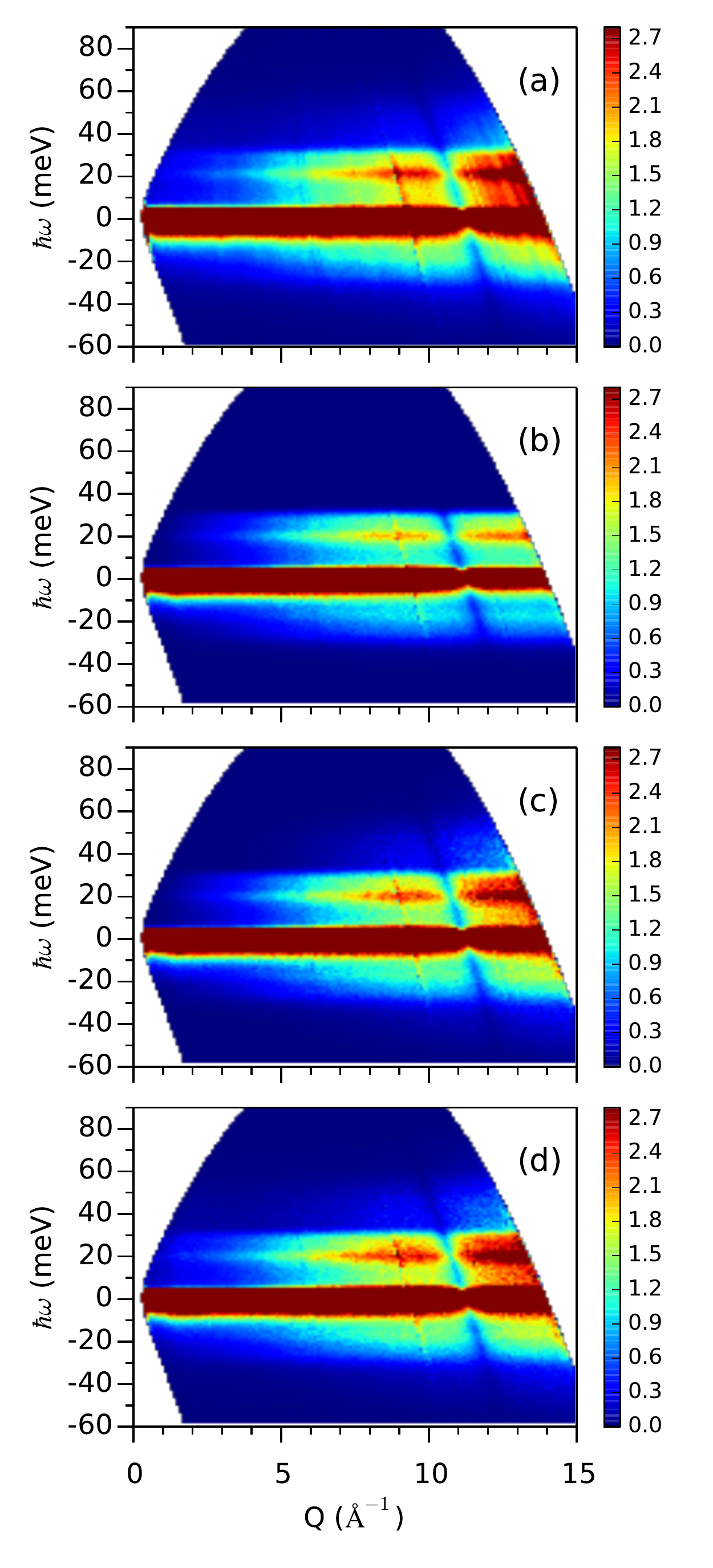}
\caption{$I(Q, \omega)$ plots of vanadium inelastic spectra obtained
  from experiment and simulations for a vanadium plate at room temperature
  in the ARCS instrument. 
  (a) Experiment
  (b) Simulation without a multi-phonon kernel. Multiple scattering
  was turned off.
  (c) Simulation with a multi-phonon kernel. Multiple scattering was
  turned off.
  (d) Simulation with a multi-phonon kernel. Multiple scattering was
  turned on.
}
\label{vanadium-iqe} 
\end{figure}

\begin{figure}
\centering
\includegraphics[width=8cm,clip]{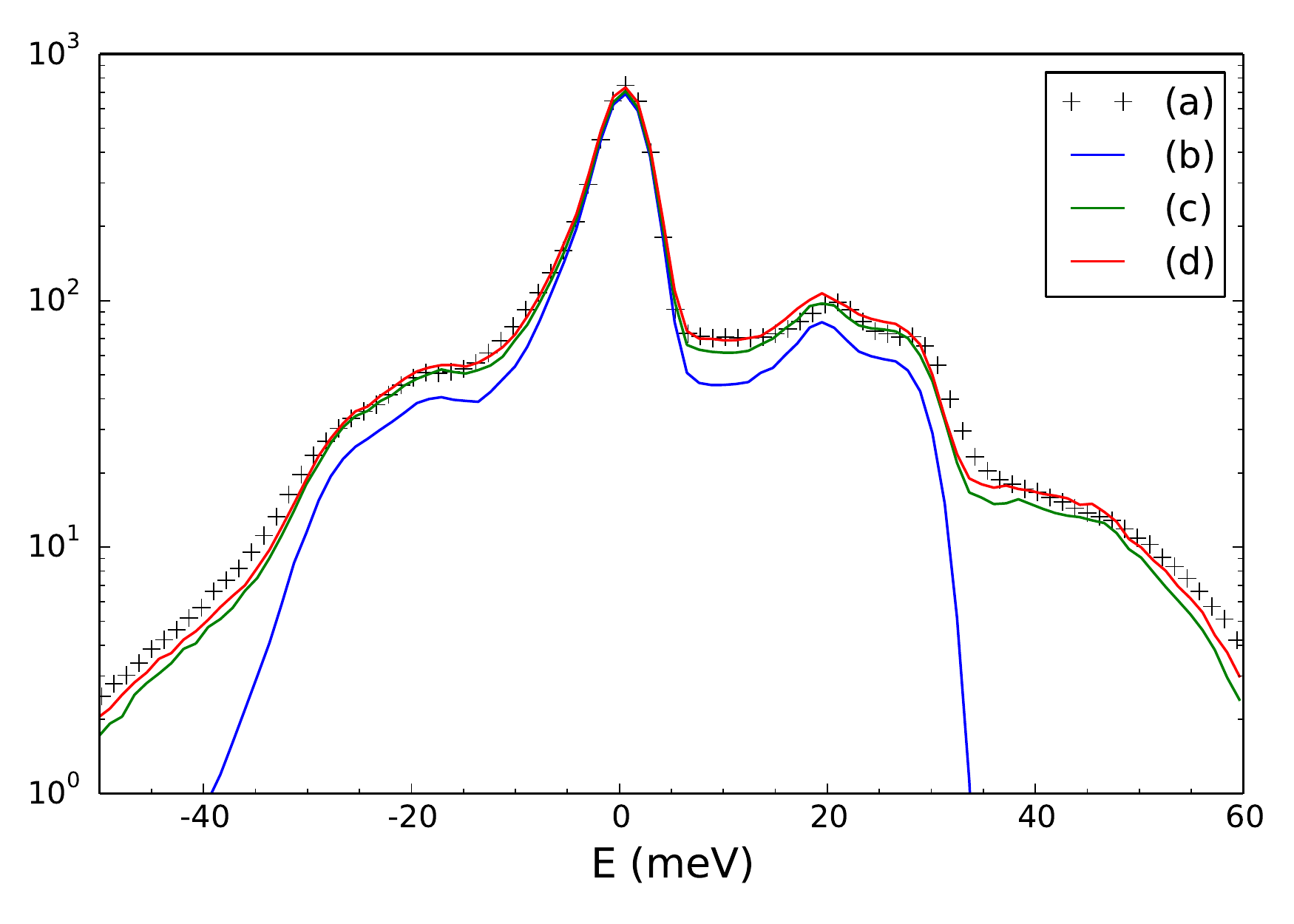}
\caption{Energy spectra integrated 
 over Q range (8,12) inverse angstrom, obtained
 from experiment and simulations for a vanadium plate placed in the
 ARCS beam. The intensity axis is in log scale. 
 (a)-(d) same as Figure~\ref{vanadium-iqe}
}
\label{vanadium-ie} 
\end{figure}
 
Here we present experimental and simulated inelastic
spectra for a
50~mm$\times$50~mm$\times$1.2mm
vanadium plate sample
in the ARCS instrument.
A quick calculation using the total scattering cross section
for V shows that such a sample is a 6\% scatterer.
Vanadium, being the regular calibration standard in neutron scattering
experiments, was chosen as the first example for its simplicity 
(incoherent scattering cross section is much larger than
that of coherent scattering, and hence the scattering is isotropic).
% incoherent neutron scattering dominates the scattering signals
In the experiment, the sample was approximately perpendicular to the beam,
and the incident energy was tuned to 117~meV using 
a Fermi chopper with 1.5~mm slit spacing and 550~mm radius of curvature,
rotating at 600~Hz.
In the simulation,
% the incident beam was at $\sim$117~meV with the Fermi chopper
% choice matching the experimenal one.
the instrument parameters used matched the experimental ones.
The simulation sample assembly contains only one homogeneous scatterer for the
vanadium plate, which was tilted 96.6~degrees from the beam direction.
Different scattering kernels were used for different simulations, but 
one incoherent elastic kernel and one single-phonon
incoherent inelastic kernel were included for all.
Incoherent scattering cross section of 5.08~barns and
absorption cross section of 5.08~barns \cite{sears1992xs} were used.
All phonon-releated scattering kernels use the phonon density of states
(DOS) calculated from a Born-von Karman (BvK) model \cite{maradudin1971theory},
which used force constants originally reported in Ref.~\cite{ColellaPRB1970},
and tabulated in the Landolt Bornstein series of
material properties~\cite{LBseries_Phonon}.
Only one universal scale factor was applied to the intensities of
all simulated spectra to match the experimental data.

Shown in Figure~\ref{vanadium-iqe} are $I(Q, \omega)$ plots.
The top panel is the experimental result. 
Panel (b) was simulated without either multi-phonon scattering or
multiple-scattering.
Panel (c) was simulated with multi-phonon scattering but
multiple-scattering was turned off.
Comparing (b) and (c), we can see how multi-phonon scattering
contributes to intensities at high $Q$ ($>\sim 10\AA^{-1}$),
especially noticeable at high energy transfer %near $E=40$~meV.
above the single-phonon cut-off energy of $\sim$35~meV.
Panel (d) was done with both multi-phonon and multiple-scattering
contributions. 
Comparing (c) and (d)
shows how the effects of the multiple scattering
are small but observable, particularly at low $Q$ values 
($<\sim 2 \AA^{-1}$).
% This is understandable because the plate
% is very thin.
As the plate is only a 6\% scatter,
one would expect the multiple scattering to be less than 0.4\%,
consistent with the observation.
It also shows that there is less $Q$-dependence in multiple-scattering
than in multi-phonon scattering.
Figure~\ref{vanadium-ie} shows the energy spectra integrated over
momentum transfer range of 8 to 12~\AA$^{-1}$ in log scale.
Similarly, we found that multi-phonon scattering contributes more
substantially than multiple scattering in this $Q$ range.

\begin{figure}
\centering
\includegraphics[width=7cm,clip]{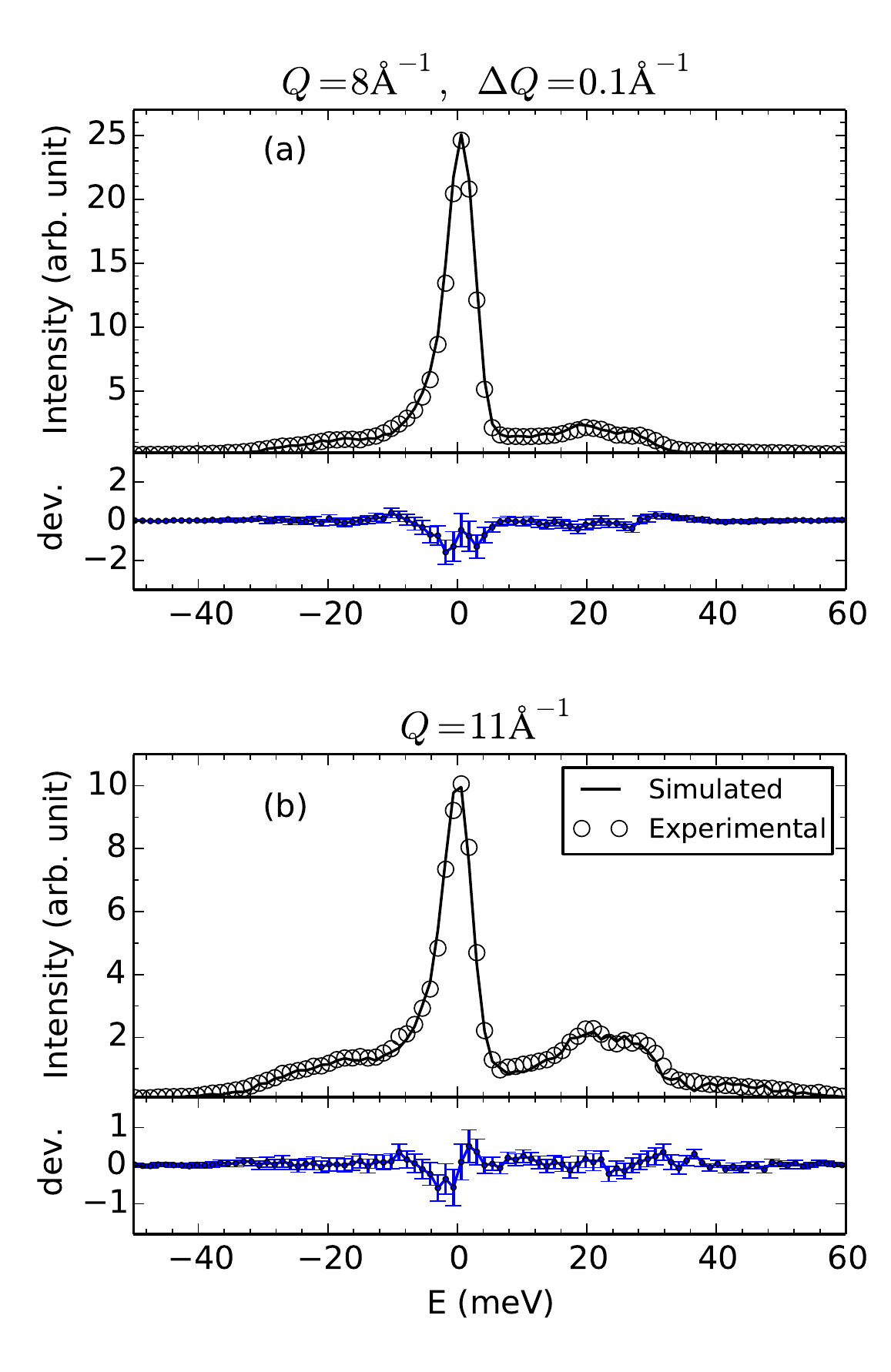}
\caption{Constant momentum cuts of
spectra as shown in Figure~\ref{vanadium-iqe}
at different $Q$ values:
(a) 8 and (b) 11~\AA$^{-1}$.
The experimental cuts were taken from 
Figure~\ref{vanadium-iqe}(a),
while the simulated cuts were taken from
the best simulation incorporating all scattering kernels
and multiple-scattering, Figure~\ref{vanadium-iqe}(d).
The spectra show decreasing 
intensities in elastic peaks and increasing relative strength
of inelastic scattering on both sides of the elastic peaks,
with larger momentum transfers.
The deviations of the simulated data from the experimental data
{($\Delta I = I_{\textrm{exp}} - I_{\textrm{sim}}$, similar thereafter)}
shown in the lower panels
are small and are within error bars.
The slightly larger deviations around elastic peaks 
may be attributed to 
slight mistmatch in determination of elastic lines,
the fact that the coherent scattering is neglected
in the simulations,
as well as uncertainties in moderator profile,
and in guide and chopper simulations.
}
\label{vanadium-ies}
\end{figure}

\begin{figure}
\centering
\includegraphics[width=10cm,clip]{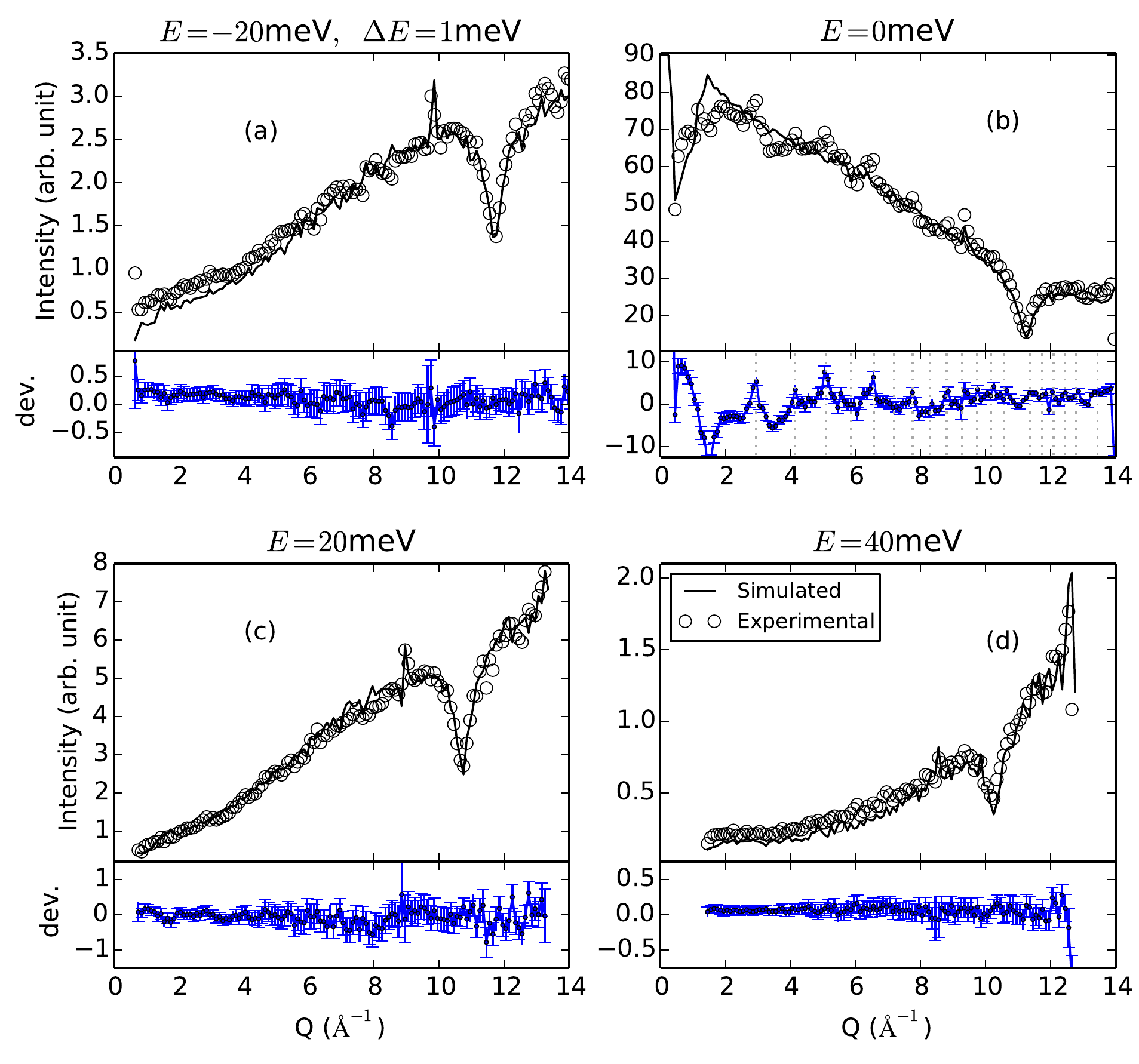}
\caption{Constant energy cuts of
spectra as shown in Figure~\ref{vanadium-iqe}
at different $E$ values:
(a) -20 (b) 0 (c) 20 and (d) 40~meV.
The experimental cuts were taken from 
Figure~\ref{vanadium-iqe}(a),
while the simulated cuts were taken from
the best simulation incorporating all scattering kernels
and multiple-scattering, Figure~\ref{vanadium-iqe}(d).
The elastic scattering in (b) weakens with higher $Q$
due to the Debye-Waller factor;
the inelastic scattering in other panels strengthens
with higher $Q$:
{for panels (a) and (c), 
this trend is 
dominated by the one-phonon scattering 
proportional to the $Q^2$ factor
in addition to the same Debye-Waller factor,
while for panel (d) for which the energy transfer of 40~meV is
beyond the single-phonon energy cut-off,
it is dominated by the two-phonon scattering
proportional to the $Q^4$ factor.}
The dip shown in every panel is typical
of the plate sample geometry and a result of 
absorption along the sample width
(the dark angle).
The deviations of the simulated data from the experimental one
(shown in the lower panels)
are in general smaller than error bars.
The peaks in the deviation curve at the lower panel of (b)
are beyond error bars, 
but they match positions of powder diffraction peaks as indicated
by the vertical dotted lines, and most likely can be accounted
for by the fact that 
the coherent scattering from V are neglected in the simulations.
}
\label{vanadium-iqs}
\end{figure}

To examine the agreement between the simulated
and the experimental spectra, 
constant momentum cuts and constant energy cuts
of the experimental data (Figure~\ref{vanadium-iqe}(a))
and the simulated data with the most complete
collection of scattering kernels
(Figure~\ref{vanadium-iqe}(d))
are presented in
Figure~\ref{vanadium-ies}
and
Figure~\ref{vanadium-iqs},
respectively.
The simulated data in general agree very
well with % the intensity profile of
the experimental data,
across the measured dynamic range
and the intensity range,
showing features such as the asymmetric line shape of 
the elastic peaks, the detailed balance,
and the Q-dependent phonon-induced energy-loss profiles
in the constant-momentum cuts,
and the dark-angle dips in the constant-energy cuts.
Also in the constant energy cuts (Figure~\ref{vanadium-iqs}), 
spikes in intensities
(e.g., at $E=-20$~meV, $Q\sim10 $~\AA$^{-1}$, and 
at $E=20$~meV, $Q\sim9 $~\AA$^{-1}$)
%resulting from a sparsity of detector pixels
%\footnote{Variation of pixel density usually
%can be accounted for in reduction software by
%normalization over solid angle coverage,
%but it can be difficult in case of abrupt variation.}
%can be observed in both the experimental 
%and simulated data.
result from insufficient filling of the displayed 
$Q$, $\omega$ histogram bins by the detector pixels.
While additional analysis may be able to correct for this effect,
it is instructive that the detailed modeling of the ARCS
detector array almost exactly matches the measured data,
while typical MC simulations would miss this effect 
due to simplified detector models.

The uncertainties of this simulation
arise from the following factors:
deviation of the simulated incident beam
from the actual neutron beam at the ARCS instrument;
% in estimating the shape of the sample
% as an ideal plate, and subsequently those
differences between the sample shape specification in simulation
and the actual sample shape, such as
its dimensions, and its orientation
relative to the incident beam;
the uncertainties in the physical and geometrical properties
of the detector packs in the detector system,
especially their positions and orientations;
the deviation of the sample temperature in simulation
from the actual temperature;
and the uncertainties in the sample material such as
impurities and contamination,
neglect of Bragg reflections in simulations,
and uncertainties in the phonon density of states (DOS) of vanadium.

The simulation of the incident beam is in part
validated by comparison to the first beam monitor,
and scattering from 
2,5-diiodothiophene (C$_4$H$_2$I$_2$S)
\cite{abernathy2012design}.
This partial validation is corroborated by
the agreement observed in the elastic peaks
of I($E$) spectra (Figure~\ref{vanadium-ies}).
The sample dimensions and its position and orientation
influence 
the position of the elastic peaks, and the position
and the width of the dark angle, which
all show excellent agreement between experimental
and simulated data.
% The dimensions of the sample also affect the magnitude
% of multiple scattering, % (MS),
% but they have to be big enough for MS to be unnegligible.
% but this has been shown to be negligible.
Uncertainties in positions and orientations of the detector packs
can result in incorrect mapping from pixels to $Q$, $E$
coordinates.
The fact that the simulated data show the same
artifacts at the low-pixel-count detector area as the experimental
data indicates that these errors are minimal.
% The decaying profile of
% the elastic I($Q$) curve (Figure~\ref{vanadium-iqs}(b))
% is governed mainly by the Debye Waller factor,
% and the detailed balance
% demonstrates itself in the I($E$) plots
% (Figure~\ref{vanadium-ies});
% both are indicators of the sample temperature,
% and both show excellent agreements between experiment and simulation.
The deviation of experimental sample temperature from
the simulation value is not expected to be larger than 10~Kelvin,
which is consistent with excellent match in detailed balance
between the simulation and the experiment shown in 
Figure~\ref{vanadium-ies}.
An examination of Figure~\ref{vanadium-iqs}(b) shows 
a series of sharp peaks, which arises from
the Bragg reflections in the sample that were not
included in the simulations.
These may account for some differences in the elastic peaks
in Figure~\ref{vanadium-ies} as well.
The excellent agreement in all the other cuts of Figure~\ref{vanadium-iqs}
shows this does not affect the spectroscopic results.
The remaining factor is the uncertainties in the phonon DOS,
which is the main source of uncertainties
in this simulation.

\begin{figure}
\centering
\includegraphics[width=8cm,clip]{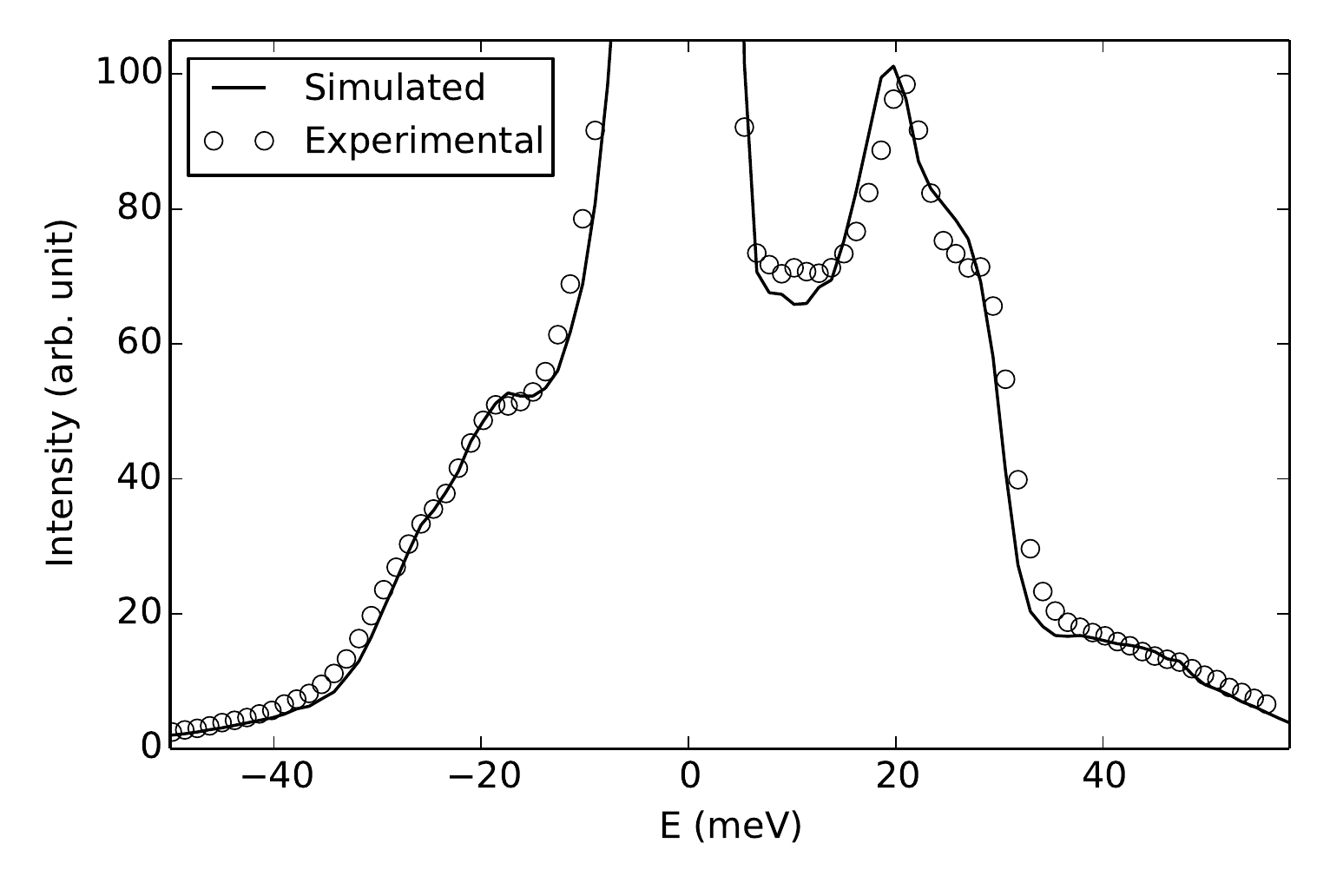}
\caption{Experimental and best-simulated energy spectra 
obtained by integrating over Q range (8,12) inverse angstrom
of the spectra as shown in 
Figure~\ref{vanadium-iqe} (a) and (d).
The intensity scale is expanded to highlight the inelastic
spectra.
}
\label{vanadium-ie-zoom}
\end{figure}

The discrepancy between the simulated data and the experimental
data, relevant to researchers interested in phonons,
is easier to observe from Figure~\ref{vanadium-ie-zoom}.
The double peaks near 20 and 28~meV both appear to shift
to the right in the experimental data, compared to
the simulated one.
This discrepancy has its origin in the phonon DOS
used in the simulation,
which is computed
from a BvK model fitted to an X-ray determination
of phonon dispersions\cite{ColellaPRB1970}.
Consistent with the discrepancies observed here,
this BvK model slightly underestimated phonon energies
when compared to a phonon DOS spectrum measured
using inelastic neutron scattering
(see Figure 10 of \cite{ColellaPRB1970}).
Actually, 
the uncertainty of measured vanadium phonon DOS reported
over the years is larger than the discrepancy
observed here (see, for example, Figure 2 of \cite{Gupta1978}).
% This result suggests that
% Fitting to I(E) spectrum or phonon DOS is a possible way to
% derive accurate force constants, parameters in the
%BvK model ... cite Lisa paper.

\subsection{Aluminum}
\label{calib-al}

% 47435

\begin{figure}
\centering
\includegraphics[width=12cm,clip]{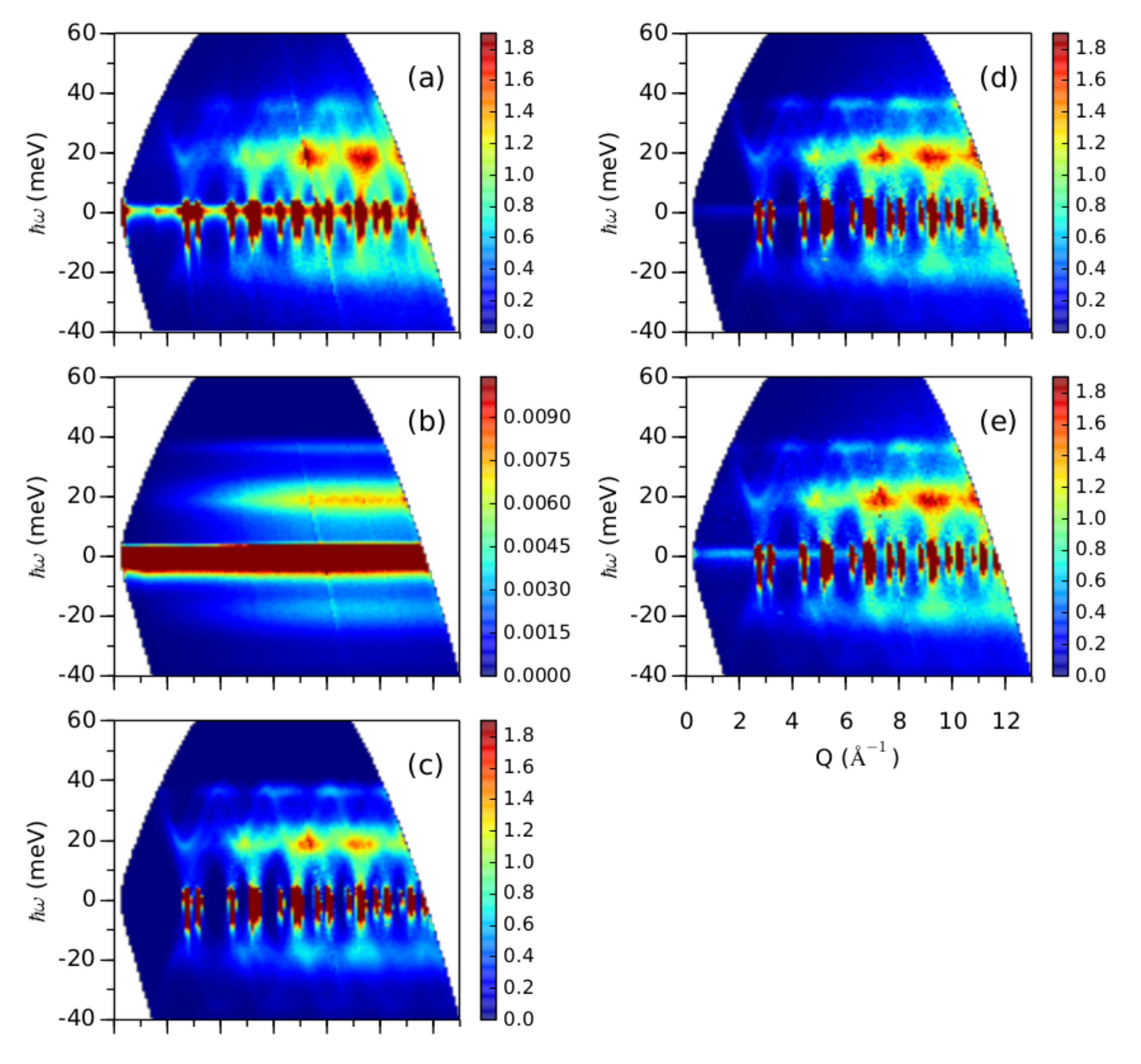}
\caption{$I(Q, \omega)$ plots of aluminum inelastic spectra obtained
  from experiment (a) and simulations (b)-(e) of an aluminum plate at 
  room temperature in
  the ARCS instrument. Scattering kernels included in the simulations
  are (b) incoherent elastic and incoherent inelastic single-phonon
  scattering (see note on intensity scaling in text), (c) coherent
  elastic (powder diffraction) and coherent inelastic single-phonon
  scattering, (d) all kernels included in (b) and (c) plus a
  multi-phonon kernel using an incoherent approximation and (e) all
  kernels included in (d) plus multiple scattering.
}
\label{aluminum-iqe} 
\end{figure}
 
\begin{figure}
\centering
\includegraphics[width=8cm,clip]{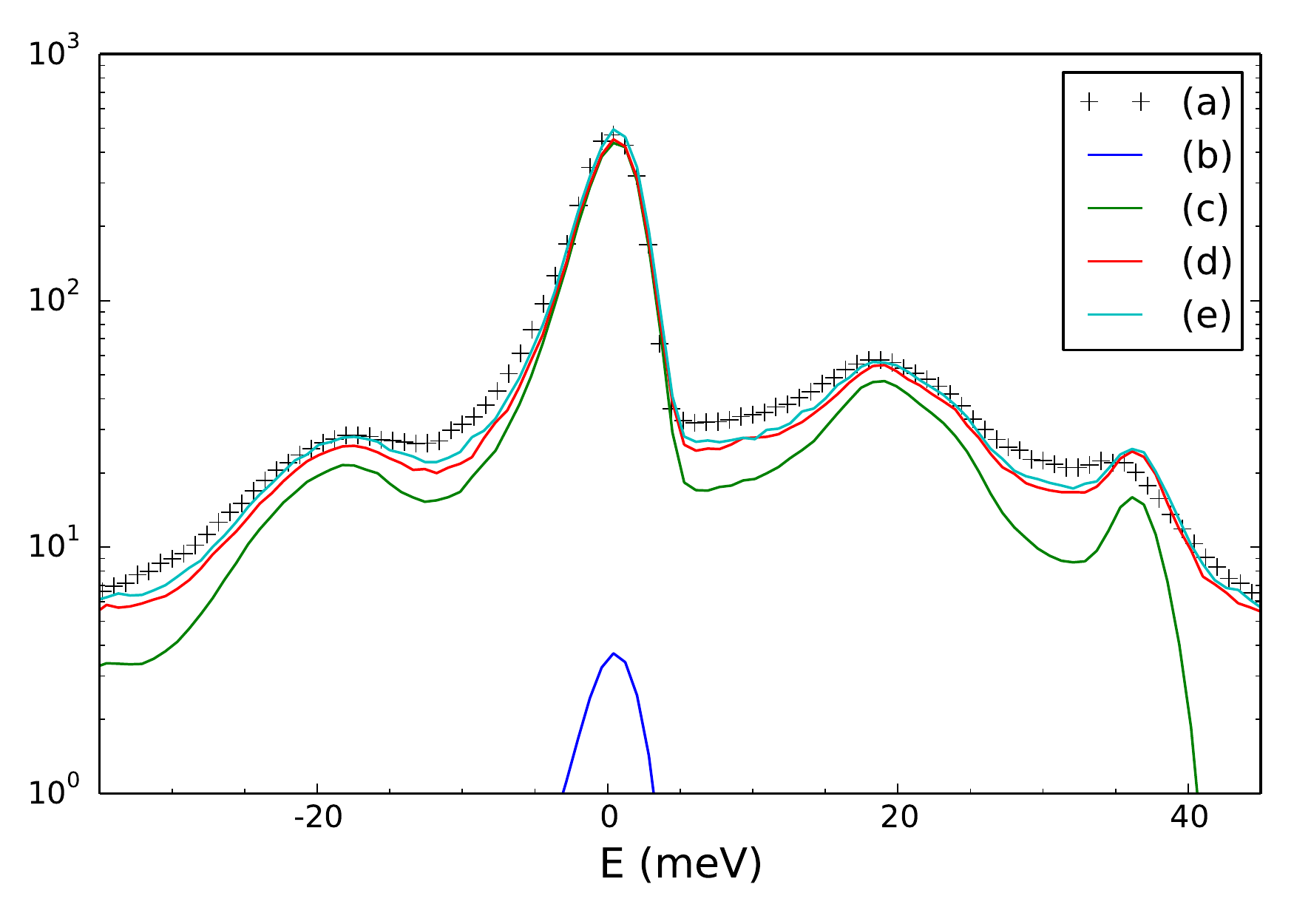}
\caption{Energy spectra integrated 
 over Q range (6,10) inverse angstrom, obtained
 from experiment and simulations for an aluminum plate placed in the
 ARCS beam. 
 The intensity axis is in log scale.
 (a)-(e) same as Figure~\ref{aluminum-iqe}
}
\label{aluminum-ie} 
\end{figure}

In this example, we present experimental and simulated inelastic
spectra from a 
60~mm$\times$60~mm$\times$4~mm
polycrystalline aluminum (1100 alloy) plate 
in the ARCS instrument.
A quick calculation using the total scattering cross section
for Al shows that such a sample is a 5\% scatterer.
In the experiment, the sample was placed approximately 
at 135~degrees from the beam,
and the incident energy was tuned to 80.5~meV using 
the same %1.5~mm slit spacing
Fermi chopper slit package as used in the previous example for vanadium,
spinning at 480~Hz.
In the simulation, the instrument parameters used
matched the experimental ones.
The simulation sample assembly contains
only one homogeneous scatterer for the
aluminum plate. 
Different simulations made use of different
combinations of scattering kernels.
Tabulated Al cross sections \cite{sears1992xs} were used.
All phonon-related scattering kernels use 
phonon energies and polarization vectors
computed on a regular grid in a Brillouin zone from a BvK model
\cite{Gilat1966Al}. %, which was derived from triple-axis neutron
%scattering data along 100, 110, and 111 directions.
The broadening of the phonon modes was not included.
%coherent inelastic one-phonon scattering kernel.
Only one universal scale factor was applied to all simulated spectra
to match the experimental data.

Figure~\ref{aluminum-iqe} shows $I(Q, \omega)$ plots for the aluminum
plate.  The experimental result is given in panel (a) and panels
(b)-(e) show simulated data.  
In (b) only the incoherent elastic and incoherent
single phonon scattering are included.  
In this plot the maximum intensity for the color scale is reduced by the ratio of 
incoherent/coherent cross sections of aluminum,  so one can see the
details in the plot.  
% Nevertheless 
The numerical values on the
colorbar give an indication of the scaling.
In (c) only the coherent elastic (powder diffraction) and the coherent
single-phonon inelastic scattering are included.  
In (d), all of the
kernels in (b) and (c) with the addition of a multi-phonon kernel
using the incoherent approximation.  
In (e), all of the kernels in (d) are
used with multiple scattering turned on.
Overall the features shown in the experimental data (a)
and the simulated data (e) agree very well.
Comparison of (b) and
(c) shows that coherent scattering gives rise to more features such as
diffraction peaks and phonon dispersion curves.
It is evident from
comparing (c) and (d) that multiphonon scattering increases in
intensity at higher $Q$.  
The most obvious difference in (d) and (e) is
in the elastic line, which shows that multiple scattering of 
coherent elastic scattering seems to
contribute similarly to incoherent scattering in the elastic line. 
The elastic lines in (a) and (e) seem to show that the sample used in the
experiment may contain traces of an additional phase, most likely from
a surface layer of Al$_2$O$_3$.
Figure~\ref{aluminum-ie} shows energy spectra integrated over
a range of momentum transfer from 6 to 10~\AA$^{-1}$ in log scale.
Similar to the vanadium measurement, we found that 
multi-phonon scattering plays a significant role in the observed intensity,
whereas multiple scattering plays little role in this $Q$ range.
The is expected as the sample is only
a 5\% scatterer.

\begin{figure}
\centering
\includegraphics[width=7cm,clip]{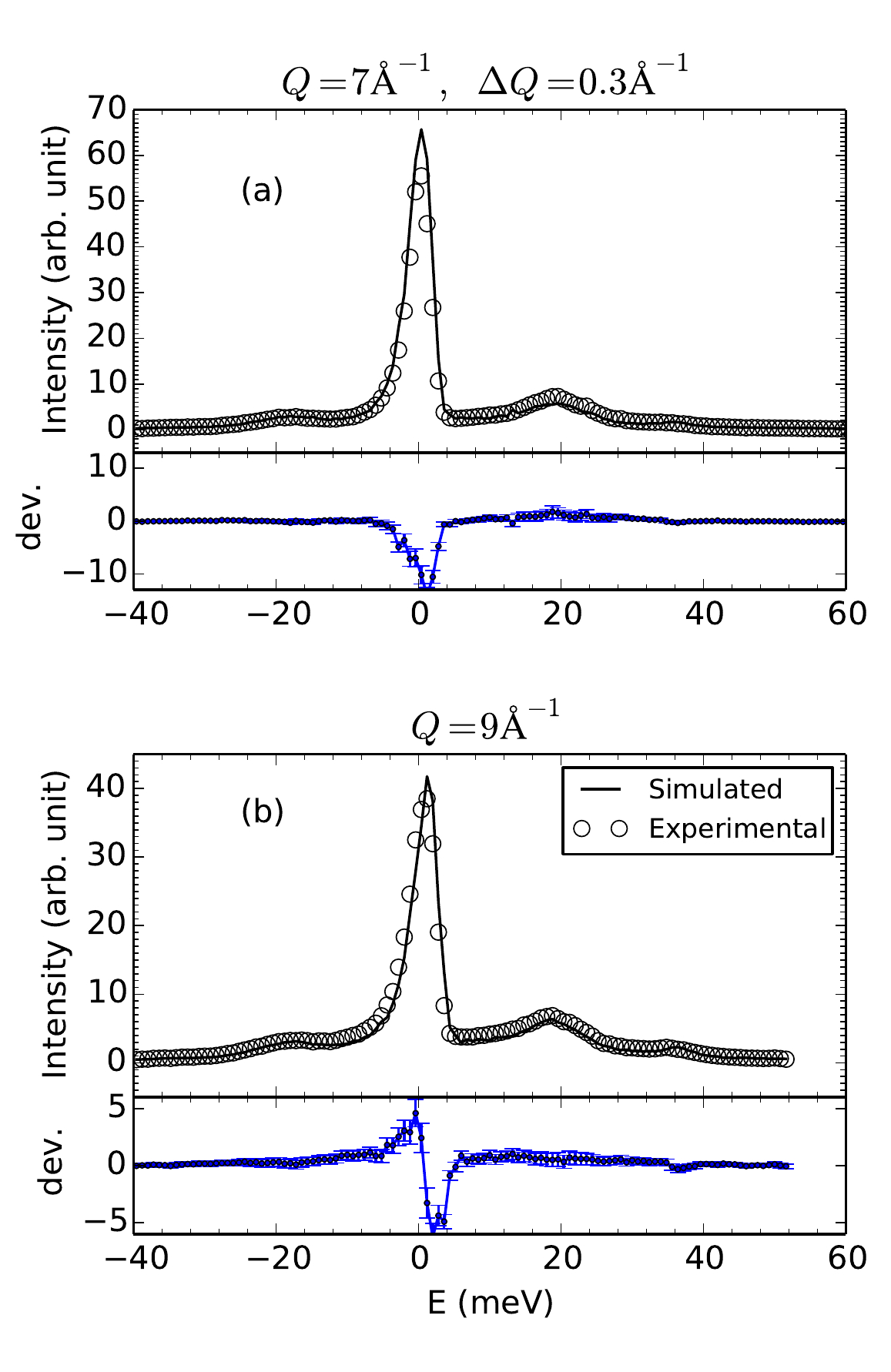}
\caption{Constant momentum cuts of
spectra as shown in Figure~\ref{aluminum-iqe}
at different $Q$ values:
(a) 3 (b) 5 (c) 7 and (d) 9 \AA$^{-1}$.
The experimental cuts were taken from 
Figure~\ref{aluminum-iqe}(a),
while the simulated cuts were taken from
the best simulation incorporating all scattering kernels
and multiple-scattering, Figure~\ref{aluminum-iqe}(e).
The deviations of the simulated data from the experimental data
are plotted in the lower panels.
The main contribution to the larger deviations around the elastic peaks 
most likely comes from the texture in the sample.
}
\label{aluminum-ies}
\end{figure}

\begin{figure}
\centering
\includegraphics[width=10cm,clip]{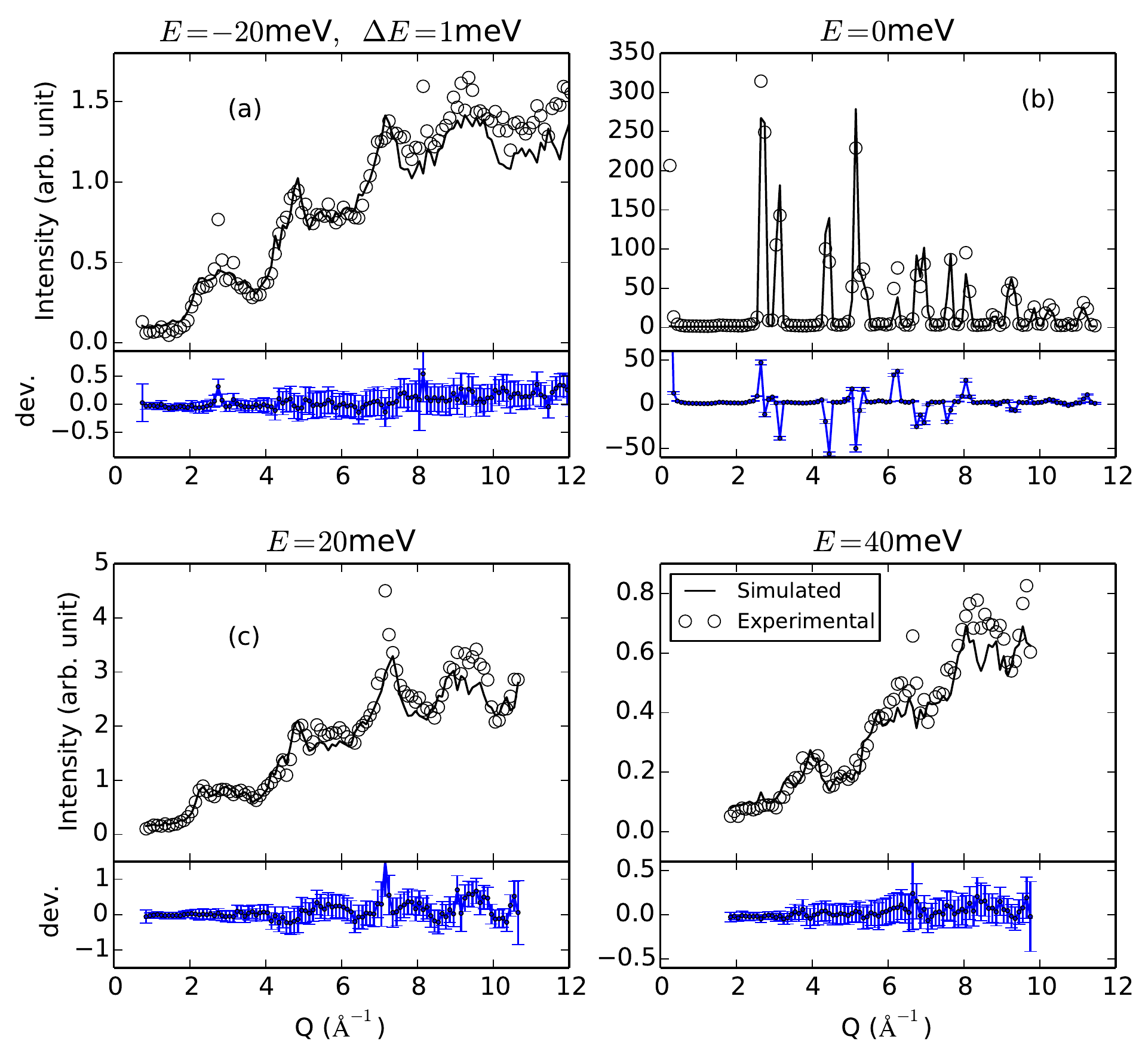}
\caption{Constant energy cuts of
spectra as shown in Figure~\ref{aluminum-iqe}
at different $E$ values:
(a) -20 (b) 0 (c) 20 and (d) 40 meV.
The experimental cuts were taken from 
Figure~\ref{aluminum-iqe}(a),
while the simulated cuts were taken from
the best simulation incorporating all scattering kernels
and multiple-scattering, Figure~\ref{aluminum-iqe}(e).
The deviations of the simulated data from the experimental one
(shown in the lower panels)
in general are smaller than the error bars.
The discrepancies in (b) are from differences
in the measured and calculated Bragg peak intensities.
These arise from the fact that the sample is not a perfect powder.
}
\label{aluminum-iqs}
\end{figure}

Similar to the vanadium study in the previous section,
we further examine the agreement between the simulated
and the experimental spectra
by taking constant momentum cuts and constant energy cuts
of the experimental data (Figure~\ref{aluminum-iqe}(a))
and the simulated data with the most complete
collection of scattering kernels
(Figure~\ref{aluminum-iqe}(e)).
These cuts are presented in
Figure~\ref{aluminum-ies}
and
Figure~\ref{aluminum-iqs},
and they again display strong agreement
between the experimental and simulated data.
%in case of $I(E)$ spectra 
%in Figure~\ref{aluminum-ies},
%and slightly worse in case of $I(Q)$ spectra
%in Figure~\ref{aluminum-iqs}.
% showing some additional features such as diffraction peaks
% (Figure~\ref{aluminum-iqs}(b)).
For Figure~\ref{aluminum-ies}, the maximum deviations
occur near the elastic peak.
Similarly, the biggest differences in Figure~\ref{aluminum-iqs}
are observed for the elastic scattering in (b).
These indicate that the measured Bragg peak intensities vary
from the model: an isotropic powder average for Al.
This could arise from texture,
preferred grain orientation,
or incomplete sampling of the powder\cite{young1993rietveld}.

The uncertainties of this simulation,
like the case of vanadium,
arise mainly from the uncertainties in the phonon data,
which is the major input to
all phonon-related kernels used in the simulation,
only in this case the phonon data is not just
a simple DOS curve.
This phonon data input, existing in the form of
energies and polarization vectors of all phonon branches,
was computed 
for all points on a regular grid 
inside the first Brillouin zone,
from a BvK model fit to 
phonon dispersions measured along 
the [100], [110], and [111]
directions\cite{Gilat1966Al}.
Small changes in the force constants in the BvK model
can result in large fluctuations of the computed 
phonon energies and polarizations.
Hence,
the discrepancy between the experimental data (Figure~\ref{aluminum-ie}(a))
and the simulated data (with all kernels, Figure~\ref{aluminum-ie}(e))
was not unexpected.
Another possible contributing factor is
the fact that
the broadening of Al phonon modes 
\cite{MKreschPRB2008Al, tang2010anharmonicity}
was not taken into account in our simulation.

\subsection{K$_2$V$_3$O$_8$ single crystal}
\label{kvo}

We now present experimental and simulated inelastic
single-crystal spectra for a K$_2$V$_3$O$_8$ sample \cite{lumsden2006prbkvo}
measured at the HYSPEC
instrument \cite{Barry2015}, a hybrid spectrometer at SNS 
with a focusing monochromator and a movable detector vessel.
The sample consisted of 5 co-aligned
crystals with an overall cylindrical shape 
approximately 3.8~cm in diameter and 
2.5~cm in height.  
The sample was oriented so that
at zero degrees of rotation angle its [100] direction was along the beam
and its [001] direction pointed upward vertically.
Measurements were performed at 1.5~K with an incident energy of 
$\sim$7~meV with a Fermi chopper frequency of 180~Hz.
The detector vessel was oriented so that neutrons scattered from the
sample in the horizontal plane were measured in the scattered angle
range of -75 to -15~degrees. 
To span reciprocal space, the goniometer angle was swept from 
40 to -50~degrees in 0.5~degree steps and then from -50 to -28~degrees
in 1~degree steps.

The simulation was done for a scan matching the experimental setup.
The sample assembly contains one homogeneous scatterer for the
K$_2$V$_3$O$_8$ sample with a shape matching the overall shape of 
the sample in the experiment.
Two scattering kernels were included in the simulation:
one incoherent elastic scattering kernel to approximate the elastic
line, and one dispersion-surface scattering kernel for simulating
the scattering from the spin-wave.

In K$_2$V$_3$O$_8$ 
the long-wavelength spin-wave dispersion is well described
by a result for the quantum (S=1/2) square lattice Heisenberg antiferromagnet:

\begin{equation}
  E(\textbf{Q}) = 2 \tilde{J} \sqrt{1-\gamma_{Q_{//}} ^2}
\end{equation}
with the dynamic structure factor
\begin{equation}
  S(\textbf{Q}, E) = \frac{1 - \gamma_{Q_{//}} } {2 \sqrt{ 1-
      \gamma^2_{Q_{//}} }} \; \delta(E-E(\textbf{Q}))
\end{equation}
where
$\gamma_{Q_{//}} = \cos(h\pi) \cos(k\pi)$, $2\tilde{J} = 2.563$ \cite{lumsden2006prbkvo}.
The differential cross section is 
\begin{equation}
  {\left(\frac{d^2\sigma}{d\Omega dE_f}\right)} \sim
  \frac{k_f}{k_i} \; \left| F(Q) \right|^2
  [n(E)+1] (1+\cos^2 \phi) S(\textbf{Q}, E) 
\end{equation}
where $k_i$ and $k_f$ 
are the magnitudes of the incident and final neutron wave vectors, 
$F(Q)$ is the magnetic form factor for V$^{4+}$ \cite{BrownMagneticFormFactor}, 
$n(E)+1$ is the Bose occupation factor,
and $1+\cos^2 \phi$ is a polarization term
($\phi$ is the angle between $\textbf{Q}$ and the easy c-axis).

\begin{figure} 
\centering
\includegraphics[width=7cm,clip]{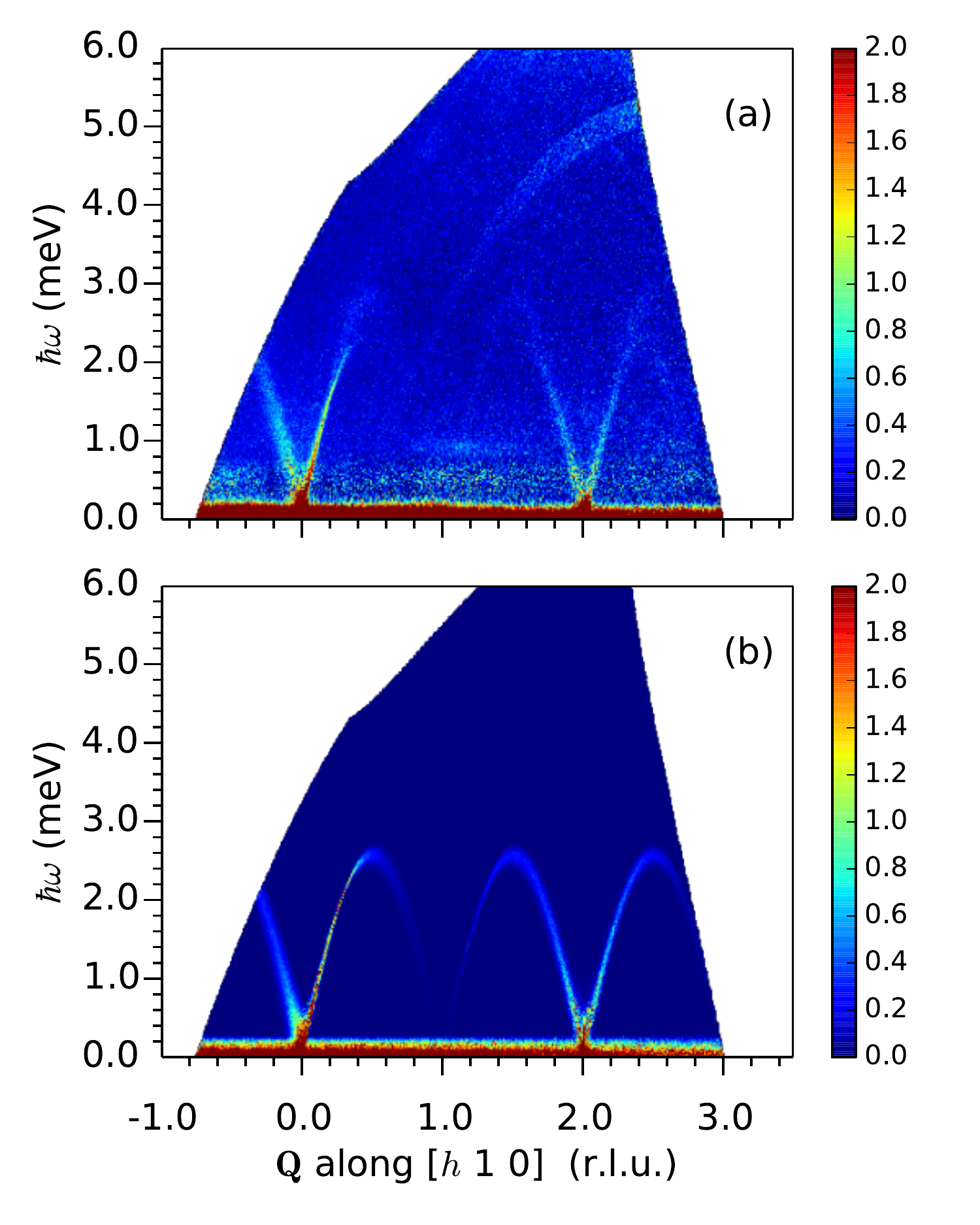}
\caption{
Slices along the $\textbf{Q}$=[$h$10] axis and the energy axis
from (a) experiment and (b) simulation,
taken in the range $l$=(-0.3, 0.3) and $k$=(0.93,1.07).
}
\label{kvo-h10-slice} 
\end{figure}

\begin{figure} 
\centering
\includegraphics[width=7cm,clip]{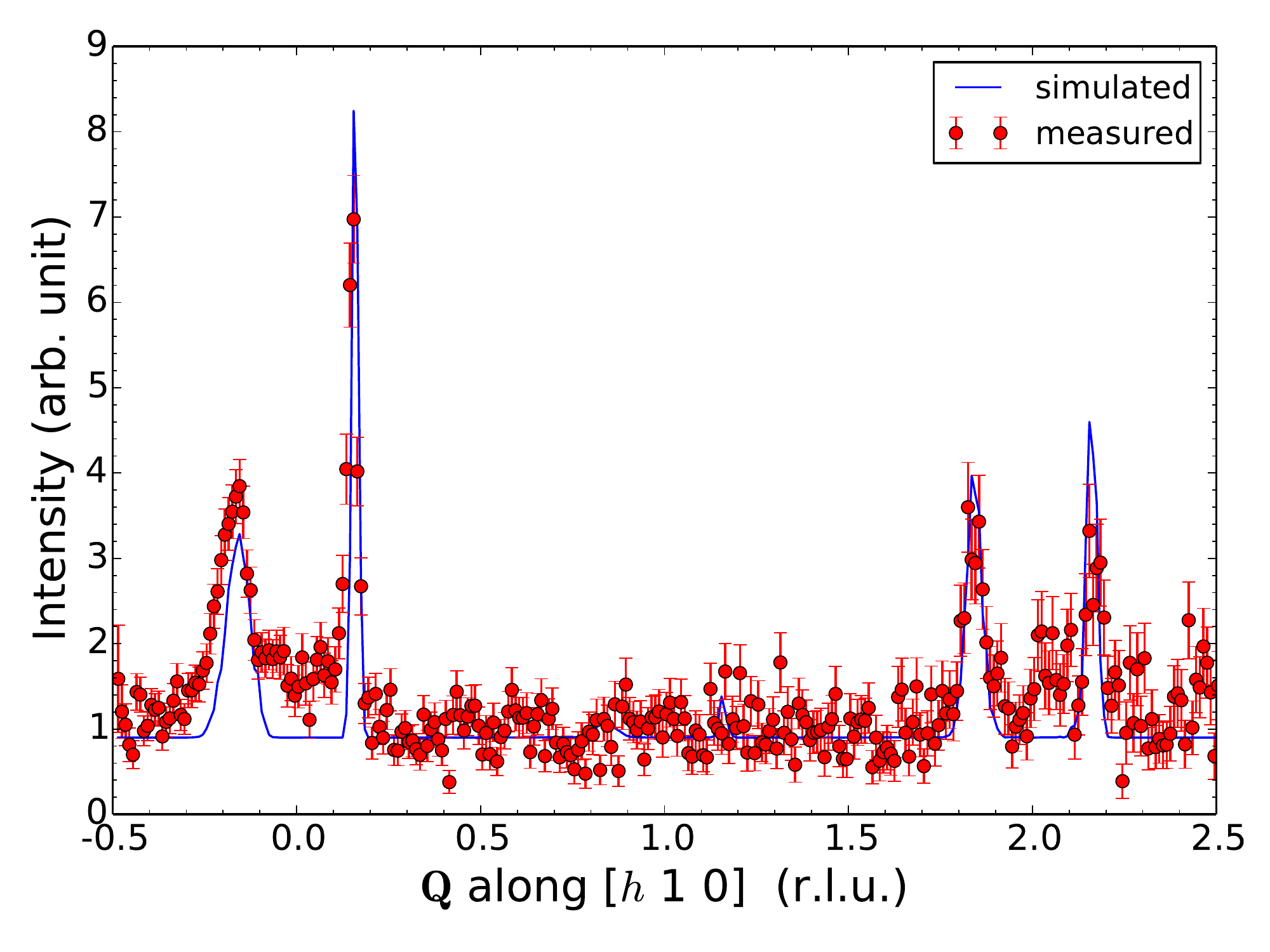}
\caption{Constant energy cuts of the slices as shown in
  Figure~\ref{kvo-h10-slice}
  in the energy range of [1.2, 1.3] meV.
  A background term and a scaling factor were the only
  parameters used to fit
  the simulated data to experimental data.
  The goodness of fit, reduced $\chi^2$, was 2.56.
}
\label{kvo-h10-cut} 
\end{figure}

In the treatment of both experimental and simulated data, Mantid was
used to reduce the measured NeXus file to NXSPE format,
a specialized intermediate inelastic neutron data format,
at each
goniometer angle, and then a Python program was used to project data 
in NXSPE files to the four dimensional $\textbf{Q},\; E$ space.
Slices along desired $\textbf{Q}$ directions could then be taken.

Shown in Figure~\ref{kvo-h10-slice} are experimental and simulated
slices along axes of $\textbf{Q}=[h10]$ and energy.
The spin wave
dispersion in the energy range between $\sim$0.3 and $\sim$2.1~meV 
shows good agreement between experimental and simulated data.
In the experimental data,
the noise in the energy range of [0.25, 0.7]
is higher than the noise at higher energies. 
This is due to  elastic scattering from the cryostat, which
was subtracted from the data.
Both plots show a distinctive feature --
for the dispersion near [010] the right branch is sharper than the
left branch,
while the dispersion branches near [210] are largely symmetric
in their broadening.
This effect is more clear in Figure~\ref{kvo-h10-cut}, where 
constant-energy cuts over [1.2,~1.3]~meV
are plotted against $\textbf{Q} = [h10]$ for 
the experimental (circles with error bars) and 
the simulated (solid line) intensities.
Both the experimental data and simulated data show 
that the peak near $h=0.15$ is much sharper than that near $h=-0.15$.
Such an effect is expected due to diffrent focusing conditions of the
resolution ellipsoid \cite{ShiraneFocusing, TobyThesis}.
The spin-wave dispersion above
$\sim$2.1~meV ($h\sim0.33$, for example) 
has weaker intensity
in the experimental data than expected from the simulated data,
which could be attributed to mode splitting
near the zone boundary \cite{lumsden2006prbkvo}.

\section{Conclusions}
\label{conclusions}
The MCViNE software package 
for Monte Carlo neutron ray tracing
simulations is based on modern object-oriented software design
that decomposes neutron scatterers into a hierarchy of
composite and elemental scatterers.
This software design elegantly solves
algorithmic problems including multiple scattering in a sample
assembly,
and ray tracing in a sophisticated detector system.
Multiple scattering and instrument resolution are included naturally
in MCViNE simulations,
and simulation of ray-tracing in a complex detector system such as
those of the ARCS and SEQUOIA instruments is straightforward.
Adding, removing, and modifying scattering kernels in the simulations
can help show the contributions of different types of
scattering events to the experimental data.  
The examples of scattering from vibrational and magnetic
excitations presented here and elsewhere\cite{linjyy2014UNprb}
demonstrate that such simulations can improve our
understanding of the underlying physics of neutron spectra.

Limitations of the MCViNE software package include
(1) the complexity of sample assembly or detector system
can be limited by available computing resources 
(memory and computation time);
(2) it currently only supports homogeneous scatterers,
and this means materials with non-uniform distribution of defects, 
for example,
can not be simulated without approximations;
(3) it has a limited but expandable libary of scattering kernels.
For example, for magnetic systems, it currently only supports
spin-wave dispersions that can be expressed as
analytical functions;
(4) MCViNE currently relies on components adapted from other MC
ray tracing packages to handle many neutron optical components
such as guides and choppers, and thereby inherits their limitations.

MCViNE % is an open source software and
is freely
available for the Linux platform. 
More details about the conditions of use and license can be found at
\href{http://danse.us/trac/MCViNE/wiki/license}
{http://danse.us/trac/MCViNE/wiki/license}.
Details on
build and installation, usage, source repository, and 
user support of MCViNE are available in the
documentation \cite{MCViNE}.
Feedback to the MCViNE developers can be provided through the
MCViNE user mailing list
\href{http://groups.google.com/group/mcvine-users}
{http://groups.google.com/group/mcvine-users}.

\section*{Acknowledgements}
The development of the MCViNE software was begun by J.Y.Y.L.
under the DANSE project supported by the NSF award
DMR-0520547.
The research on simulations of experiments in the ARCS, SEQUOIA, and HYSPEC
instruments was supported by the U.S. Department of Energy, Office
of Basic Energy Sciences.
G.E.G., A.A.A., D.L.A., M.D.L., B.A. were fully supported,  
J.Y.Y.L. and H.L.S. partially supported by the Scientific User Facilities
Division. 
We thank M. E. Hagen, A. Payzant, and P. Willendrup 
for stimulating discussions.
We also thank L. Li and A. Dementsov for developing the powder
diffraction scattering kernel for MCViNE, 
A. Fang for building MCViNE
adaptations of some McStas components, 
and M. Reuter and S. Campbell for updating the MANTID code
to read in the Monte Carlo generated data.

\bibliography{mcvine}

\end{document}

\begin{frontmatter}
\title{Supplemental Material for \\
  MCViNE -- An object oriented Monte Carlo neutron ray tracing
  simulation package 
}

\author[cacrcaltech,aphmscaltech,ndavornl]{Jiao Y. Y. Lin \corref{jiao}}
\ead{linjiao@ornl.gov, linjiao@caltech.edu}

\author[aphmscaltech]{Hillary L. Smith}

\author[ndavornl]{Garrett E. Granroth \corref{garrett}}
\ead{granrothge@ornl.gov}

\author[qcmornl]{Douglas~L.~Abernathy}
\author[qcmornl]{Mark D. Lumsden}
\author[qcmornl]{Barry Winn}
\author[qcmornl]{Adam A. Aczel}
\author[cacrcaltech]{Michael~Aivazis}
\author[aphmscaltech]{Brent Fultz \corref{btf}}
\ead{btf@caltech.edu}

\cortext[jiao,garrett,btf]{Corresponding author}

\address[cacrcaltech]{
  Caltech Center for Advanced Computing Research, California Institute
  of Technology
}
\address[aphmscaltech]{
  Department of Applied Physics and Materials Science, California
  Institute of Technology
}
\address[ndavornl]{
  Neutron Data Analysis and Visualization Division, Oak Ridge National Laboratory
}
\address[qcmornl]{
  Quantum Condensed Matter Division, Oak Ridge National Laboratory
}

\end{frontmatter}

%
% \linenumbers

\lstset{
  language=Python, 
  basicstyle=\ttfamily\footnotesize, 
  keywordstyle=\color{blue},
  commentstyle=\color{Gray},
  stringstyle=\color{red},
  showstringspaces=false,
  identifierstyle=\color{black}
}

\section{Why Object Oriented?}
Imperative programming is the predominant programming paradigm
employed in MC neutron ray tracing programs.
In imperative programming, 
different aspects of programming data and logic are %often 
tightly coupled.
Object-oriented programming (OOP) 
is less used in scientific programming.
Fundamental OOP concepts include
encapsulation (packing of data and associated functions), 
polymorphism (processing of objects according to type), 
dynamic dispatch (dynamically deciding which implementation to choose for an operation), 
and delegation (allowing a different object or method to do the work).
These methods promote abstraction and decoupling of data
and programming logic, %, and concerns,
generally resulting in codes that are more extensible, more
maintainable, and easier to debug 
\cite{GammaDP,MeyersEffectiveCpp,MeyersMoreEffectiveCpp,LakosLargeScaleCppSoftwareDesign}.
It should be noted that runtime efficiency of a program is not the
main focus of the OOP paradigm:
to achieve polymorphism, for
example, additional machine codes are generated by compilers to
facilitate the dynamic dispatch by using virtual tables.
OOP codes are also harder to optimize by the compiler.
It is not uncommon for OOP codes to run 
slower than programs written
with imperative programming.
However, this inefficiency can be overcome
by tremendous improvements in both developer
productivity and software sustainability.
% In the following we will show how some OOP concepts and design
% patterns are used in MCViNE and produce an extensible simulation framework.

\section{Layered Architecture of MCViNE}
\begin{figure}
\centering
\includegraphics[width=8.3cm,clip]{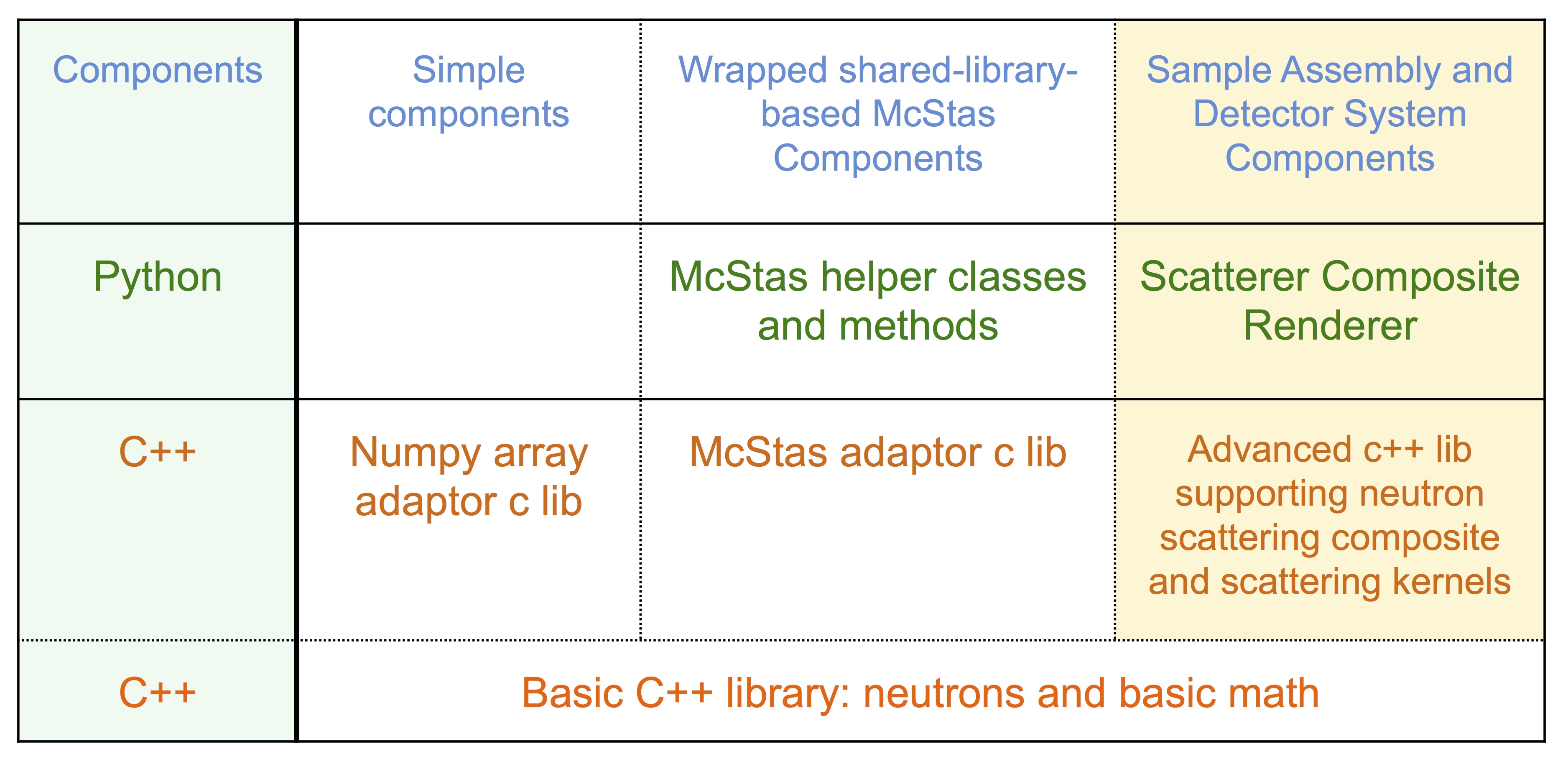}
\caption{The MCViNE architecture is divided into three categories for
  serving different neutron ray-tracing components:  simple
  NumPy-based components, McStas-based components, and components
  built with OOP principles for sample assemblies and detector
  systems.  These categories are served by the vertical integration of
  three layers comprised of Python components at the top, followed by
  python libraries, then C++ libraries.  The foundation of the MCViNE
  architecture is a C++ library with basic neutron objects and math
  functions. 
}
\label{fig-mcvine-arch} 
\end{figure}

Figure~\ref{fig-mcvine-arch} depicts the architecture of the MCViNE
package.
Two layers of libraries exist in C++.  The bottom layer is a basic C++ library that defines neutrons and some
basic math facilities such as 3D vectors.
Above this layer are three different types of C++ libraries:
1) A NumPy\cite{numpy} array adapter that allows a group of neutrons to be
manipulated easily as a NumPy array at the Python level.
2) A McStas adapter C library that provides support for MCViNE-wrapped
McStas components.
3) An advanced C++ library that supports the concept of neutron scattering
composite and scattering kernels,
a unique feature of MCViNE.
Above the three C/C++ libraries, a layer of Python helper classes and methods
are implemented to provide a Python interface to the underlying C++
functionalities.

Three types of Python components sit on top of the Python adaptation layer, including:
1) NumPy based neutron components in which a group of neutron
packets can be manipulated as a NumPy array. 
This makes it extremely easy to create
simple neutron components, and to create prototypes of more sophisticated
neutron components.
2) MCViNE-wrapped, shared-library-based McStas components 
in which a MCViNE facility automatically takes McStas components, 
compiles them into shared libraries, and binds them to Python.   
Unlike simulations in McStas, MCViNE components are not 
compiled to a monolithic executable. 
3) Generic scatterer components that support
composite scatterers and scattering kernels, 
which are described in detail in the main text. 
% Currently two such components exist, one for
% sample assemblies and one for detector systems.
% The unique feature of MCViNE, the ``composite scatterer'',
% will be explained in detail in section \ref{composite-scatterer}.

\section{\label{app:MS_algo} Multiple Scattering Algorithm}
To implement the multiple scattering algorithm in MCViNE,
the CompositeScatterer has a method ``scatterM''. 
% which is the main method for dealing with multiple scattering.
It takes one neutron as input, and outputs all
scattered neutrons, which includes the original neutron with its
probability lowered by attenuation.
The pseudo-code (in a python-like syntax) of scatterM is

\begin{lstlisting}[firstline=1, lastline=40]
def scatterM(scatterer, neutron):
  # initialize array of neutrons to be scattered
  toscatter = [neutron] 
  nms = 0 # multiple scattering order
  out = [] # initialize array of scattered neutrons
  # main loop
  while toscatter is not empty and nms < scatterer.max_ms:
    # init array of scattered neutrons
    scattered = []
    # inner loop over all neutrons that need scattering
    for neutron in toscatter:
      # if neutron probability is low, skip
      if neutron.probability < scatterer.min_prob:
        continue
      # if the neutron is leaving the scatterer, save it
      if neutron is exiting:
        out.append(neutron)
        continue
      # call interactM_path1 to compute all neutrons scattered
      # from the first encouter of the neutron with the scatterer
      scattered += scatterer.interactM_path1(neutron)
      continue
    # scattered neutrons may need to be scattered again
    toscatter = scattered
    # increment multiple scattering order
    nms += 1
    continue
  # propagate remaining neutrons
  if toscatter is not empty:
    out += propagat_out_with_attenuation(toscatter, scatterer)
  # return scattered neutrons
  return out







\end{lstlisting}
This code contains two loops. 
The outer loop (the {\it while} loop) keeps track of  two variables,
``toscatter'' (the neutrons to be scattered), and 
``nms'' (the multiple scattering order).
For every neutron in the ``toscatter'' variable
whose forward path intersects the scatterer, the
inner loop computes a set of scattered neutrons
by calling the method interactM\_path1.
%The inner loop computes for each input neutron that needs scattering 
%(its forward
%path intersects the scatterer) a set of scattered neutrons. 
This set of scattered neutrons 
(``path1-scattered-neutrons'' for future reference)
may be scattered zero, one, or more
times by the scatterer, but each of them has a (multiple-scattering)
path that is always inside the first constituent scatterer that it
encounters (``scatterer1'').  
There can be only one exception: if the original position of the 
neutron is outside the scatterer, the path segment that leads the
neutron from its original position to
its entry point of scatterer1 can be in vacuum or air.
The computed neutrons are accumulated into the variable
``scattered''
which later is renamed to ``toscatter''
for the next round of this inner loop.
The inner loop skips over neutrons that have probabilities lower than
a threshold selected by the user.
The outer loop keeps running the inner loop until either all the
neutrons scattered in the inner loop leave the scatterer (no
more interceptions), or the maximum multiple scattering order, 
as set by the user, is reached.

The ``interactM\_path1'' method of a composite scatterer takes one neutron as input
and computes ``path1-scattered-neutrons''.
Its implementation finds the first scatterer ``scatterer1'' that intercepts
the neutron, transforms the neutron to the coordinate
system of scatterer1, calls the 
``interactM\_path1'' method of  scatterer1, 
and finally transforms the scattered neutrons
back into the coordinate system of the host scatterer.
Scatterer1 could be a composite scatterer itself, 
in which case the call of the ``interactM\_path1'' method
would recurse into itself.
If scatterer1 is an elemental scatterer, the elemental
scatterer's interactM\_path1 method will be called,
hence stopping the recursion.
The selection of the interactM\_path1 method is dynamically done by
C++, thanks to the polymorphism.

The only elemental scatterer implemented so far is the HomogeneousScatterer.
Its implementation includes an interactM\_path1 method that
can be illustrated by the following pseudo code:

\begin{lstlisting}[firstline=50, lastline=100]

def interactM_path1(hs, neutron):
  # inputs:
  #   hs: homogeneous scatterer
  #   neutron: input neutron
  # output:
  #   scattered neutrons. these neutrons are scattered zero, one, or
  #   more times by "hs", but the paths of those neutrons are always
  #   inside "hs" except the leading path to the entry point.
  #   
  # initialize the neutron to be scattered
  toscatter = neutron
  nms = 0 # multiple scattering order
  out = [] # initialize array of scattered neutrons
  # main loop
  while toscatter is not null and nms < hs.max_ms:
    # if neutron probability is low, skip
    if toscatter.probability < scatterer.min_prob:
      toscatter = null
      break
    # let neutron interact with the scatterer once
    transmitted, scattered = hs.interactM1(toscatter)
    # transmitted neutron should be saved
    out.append(transmitted)
    # if the neutron is not inside this scatterer
    # done.
    if not isinside(scattered, hs):
      out.append(scattered)
      toscatter = null
      break
    toscatter = scattered
    # increment multiple scattering order
    nms += 1
    continue
  # remaining neutrons
  if toscatter is not null:
    out.append(toscatter)
  # return scattered neutrons
  return out
\end{lstlisting}
The pseudo-code is mostly self-explanatory, and the only thing that
deserves explanation is the
method interactM1 of a HomogeneousScatterer.
This method takes one neutron as input, and computes two output
neutrons, transmitted and scattered.
The transmitted neutron is the forward-propagated neutron with its probability
lowered by attenuation along its first segment of its forward path through
the scatterer.  
The scattered neutron is computed by randomly picking one point along 
the first segment of the forward path of the 
neutron, propagating the neutron to that point with attenuation,
selecting a kernel and calling the kernel's scatter method to change
the state of the neutron, and finally adjusting the probability of the neutron.

\section{Scattering kernels used in the examples}

{\bf Incoherent elastic scattering} is simply given by \cite{squires2012introduction}
\begin{equation}
  \frac{d\sigma}{d\Omega} = \frac{\sigma_{\rm inc}}{4\pi} N \exp(-2W)
\end{equation}
where $\exp(-2W)$ is the Debye-Waller factor, and $\sigma_{\rm inc}$ is the
incoherent cross section.

{\bf Coherent elastic scattering from powder sample}. The total cross
section for scattering neutrons into a Debye-Scherrer cone is
\cite{squires2012introduction}

\begin{align*}
 \sigma_{\tau}
  = & \frac{N}{v_0} \frac{\lambda^3}{4 \sin\frac{\theta}{2}}
  \sum_{\mathbf{\tau\prime}=\tau} |F_N(\mathbf{\tau\prime})|^2
\end{align*}
The sum is over reciprocal lattice vectors of the same length.
$\lambda$ is the wavelength of the neutron,
$\theta$ is the scattering angle, 
and $F_N$ is the nuclear structure factor.

{\bf Incoherent inelastic single-phonon scattering } is given by
\cite{squires2012introduction}
\begin{equation}
  \left( \frac{d^2\sigma}{d\Omega dE_f} \right)  _{\pm 1} = 
  \frac{\sigma_{inc}}{4\pi} \frac{k_f}{k_i} N \; \frac{\hbar^2Q^2}{2M} \exp(-2W) 
  \frac{Z(E)}{E} 
  \left \{ \coth \left( \frac{\hbar \omega}{2k_B T} \right) \pm 1 \right \}/2
\end{equation}
where 
%$exp(-2W)$ is the Debye-Waller factor,
% $\sigma_{inc}$ is the incoherent cross section, 
$k_i$ and $k_f$ are the momentum of the incident and outgoing neutrons,
respectively,
$Q$ is the momentum transfer, $E$ is the energy transfer,
$\left\{ \coth \left(\frac{\hbar \omega}{2k_B T} \right) \pm 1 \right\}/2$ is the thermal factor,
and 
$Z(E)$ is the phonon density of states as a function of phonon
energy $E$, which is the main input of the kernel.

{\bf Coherent inelastic single-phonon scattering from a powder sample} 
is given by
\cite{squires2012introduction}
\begin{align*}
  {\left(\frac{d^2\sigma}{d\Omega dE_f}\right)}_{inc\pm 1} = &
  \frac{\sigma_{coh}}{4\pi} \frac{k_f}{k_i} \frac{(2\pi)^3}{v_0} 
  \exp(-2W)  \\
  & \times \sum_s \sum_{\mathbf{\tau}} 
  \frac{ \hbar^2 ( \mathbf{Q} \cdot \mathbf{e}_s )^2 }{2M\;E_s} 
  \frac{1}{2} \left\{ \coth\left(\frac{\hbar \omega}{2k_B T}\right) \pm 1 \right\}
  \delta(E- E_s) \; \delta(\mathbf{Q} - \mathbf{q} - \mathbf{\tau})
\end{align*}
where 
$\mathbf{e}_s$ is the polarization of the phonon mode,
$E_s$ is the energy of the phonon mode, and
$\mathbf{\tau}$ is a reciprocal lattice vector.
The total cross section for a phonon mode 
of energy $E$ at $\mathbf{Q}$ can be
deduced as
\begin{equation}
  \sigma_{\mathbf{Q}} = 
   \frac{\sigma_{coh}}{4\pi} 
   \frac{k_f}{k_i}
   \frac{\left( 2\pi \right)^3}{v_0}
   \exp(-2W)
   \frac{\hbar^2 ({\mathbf{Q}}\cdot {\mathbf{e}})^2}{2M E} 
   \frac{1}{2} 
   \left\{ coth\left(\frac{\hbar \omega}{2k_B T}\right) \pm 1 \right \}
   \frac{1}{2 k_i k_f Q}
\end{equation}
The main input for this kernel is the energies and polarization vectors
of phonon modes in a Brillouin zone. 

{\bf Multi-phonon scattering} is given by \cite{sears1995phonon}

\begin{align*}
 {\left(\frac{d^2\sigma}{d\Omega dE_f}\right)} &=
 \frac{\sigma}{4\pi} \frac{k_f}{k_i} N \; S(Q,E) \\
 S(Q,E) &= \sum^{\rm {max}\;N}_{n=2} S_n(Q) \; A_n(E)
\end{align*}
where
\begin{align*}
  S_n(Q) &= \frac{(2W)^n}{n!} \exp(-2W) \\
  A_n  &= A_1 \ast A_{n-1} \\
  A_1(E) &= \frac{Z(E)} {E \gamma_0} \frac{1} {{\rm e}^{E/k_B T}-1} \\
  \gamma_0 &= \int^{E_max}_0 \coth(E/2k_B T) \frac{Z(E)} {E} dE
  % \gamma_0 &= \int^\infty_0 coth(E/2k_B T) \frac{Z(E)} {E} dE
\end{align*}
The main input is the phonon DOS.

{\bf Scattering from a dispersion surface $E=E(\textbf{Q})$ with a
 scattering function of $S=S(\textbf{Q})$} is given by
\begin{align*}
 {\left(\frac{d^2\sigma}{d\Omega dE_f}\right)} &=
 \frac{\sigma}{4\pi} \frac{k_f}{k_i} N S(\textbf{Q},E) \\
 S(\textbf{Q},E) &= S(\textbf{Q}) \; \delta(E-E(\textbf{Q}))
\end{align*}
The main inputs are analytical functions $E(\textbf{Q})$ and $S(\textbf{Q})$ as strings.

Here is an example of an xml specification for this type of kernel:
\begin{lstlisting}
    <E_vQ_Kernel 
       E_Q="2.5*sqrt(1-(cos(4.4*Qx)*cos(4.4*Qz))^2)"
       S_Q="1"
    />
\end{lstlisting}

\bibliography{mcvine}